%% file: MBF_FDG_v8.tex
\documentclass[12pt]{iopart}
\usepackage{iopams}  
\usepackage{color}
\usepackage{graphicx}
\usepackage{subfigure}
\usepackage{bm}
\usepackage{xspace}
\usepackage{times}
\usepackage{cite}
\usepackage{multirow}

\include{myshortcuts}

\newcommand{\mbf}{\mathrm{MBF}}

\newcommand{\fdg}{\mathrm{FDG}}
\newcommand{\rb}{\mathrm{RB}}
\newcommand{\txtc}[1]{\textcolor{black}{#1}}
\newcommand{\txtb}[1]{\textcolor{black}{#1}}
\newcommand{\adj}{\mathrm{adj}}
\newcommand{\glu}{\mathrm{glu}}
\newcommand{\mrm}{\mathrm{m}}
\newcommand{\mm}{M\!M}
\newcommand{\pbr}{\mathrm{PBR}}
\newcommand{\idif}{\mathrm{IDIF}}

\begin{document}

\title[Myocardial FDG Flow]{Multiparametric Cardiac $^{18}$F-FDG PET in Humans: {Pilot} Comparison of FDG Delivery Rate with $^{82}$Rb Myocardial Blood Flow}
\author{Yang Zuo$^{1}$, Javier E. López$^{2}$, Thomas W. Smith$^{2}$, Cameron C. Foster$^{1}$, Richard E. Carson$^3$, Ramsey D. Badawi$^{1,4}$, Guobao Wang$^{1}$}
\address{1. Department of Radiology, University of California Davis Medical Center, CA 95817, United States}
\address{2. Department of Internal Medicine, University of California Davis Medical Center, CA 95817, United States}
\address{{3.  Department of Radiology and Biomedical Imaging, Yale University, New Haven, CT 06520, United States}}
\address{4. Department of Biomedical Engineering, University of California at Davis, United States}
\ead{gbwang@ucdavis.edu}

\begin{abstract}
{Myocardial blood flow (MBF) and flow reserve are usually quantified in the clinic with positron emission tomography (PET) using a perfusion-specific radiotracer (e.g. $^{82}$Rb-chloride). However, the clinical accessibility of existing perfusion tracers remains limited. Meanwhile, $^{18}$F-fluorodeoxyglucose (FDG) is a commonly used radiotracer for PET metabolic imaging without similar limitations. In this paper, we explore the potential of $^{18}$F-FDG for myocardial perfusion imaging by comparing the myocardial FDG delivery rate $K_1$ with MBF as determined by dynamic $^{82}$Rb PET in fourteen human subjects with heart disease. Two sets of FDG $K_1$ were derived from one-hour dynamic FDG scans. One was the original FDG $K_1$ estimates and the other was the corresponding $K_1$ values that were linearly normalized for blood glucose levels. A generalized Renkin-Crone model was used to fit FDG $K_1$ with Rb MBF, which then allowed for a nonlinear extraction fraction correction for converting FDG $K_1$ to MBF. The linear correlation between FDG-derived MBF and Rb MBF was moderate (r=0.79) before the glucose normalization and became much improved (r$>$0.9) after glucose normalization.  The extraction fraction of FDG was also similar to that of Rb-chloride in the myocardium. The results from this pilot study suggest that dynamic cardiac FDG-PET with tracer kinetic modeling has the potential to provide MBF in addition to its conventional use for metabolic imaging. }
\end{abstract}

\section{Introduction}

Myocardial perfusion (blood flow) imaging with positron emission tomography (PET) has been applied in clinical cardiology to diagnose and characterize cardiovascular diseases \cite{Kaufmann2005, DiCarli2007, Schindler2010, Murthy2018}. Various perfusion radiotracers {(e.g., $^{15}$O-water, $^{13}$N-ammonia, $^{82}$Rb-chloride, $^{11}$C-acetate)} have been used \cite{Maddahi2014}. Despite their potential, the accessibility to these flow tracers remains limited for clinical use. For example,  
$^{15}$O-water is the gold standard for measuring blood flow \cite{Iida1988, Danad2014} but its half-life is very short (2.05 minutes), requiring onsite cyclotron for tracer production and is not approved for routine clinical use.  $^{13}$N-ammonia \cite{Muzik1993, Slomka2012} and $^{82}$Rb-chloride \cite{Mullani1983, Lortie2007, ElFakhri2009, Nesterov2014}  are the two blood flow radiotracers routinely used in clinical practice \cite{Maddahi2014}. However, $^{13}$N-ammonia also requires an onsite or nearby cyclotron due to its short half-life of 10 minutes. $^{82}$Rb-chloride can be produced by a mobile generator despite its short half-life (76 seconds). Nevertheless, the cost of a rubidium generator is $\geq\$30,000$ for every 4-6 weeks; this is only affordable to those hospitals or centers with a high throughput of cardiac patients \cite{Maddahi2012, DiCarli2007}. {$^{11}$C-acetate is another promising tracer \cite{vandenHoff2001,  Sciacca2001,  Timmer2010} but its 20-minute half-life still requires a nearby cyclotron.} Together, these practical challenges limit the access to perfusion imaging by PET.
  
$^{18}$F-fluorodeoxyglucose (FDG), which has a longer half-life of 110 minutes, is the most broadly used clinical PET radiotracer, mainly for metabolic imaging \cite{Maddahi2012}. In clinical cardiology, $^{18}$F-FDG PET is commonly used in combination with a short half-life flow tracer to evaluate flow-metabolism mismatch in the myocardium \cite{Abraham2010}. Such a two-tracer PET method is the gold standard for assessing myocardial viability \cite{Camici2008} and inflammatory conditions such as cardiac sarcoidosis \cite{Yama2003}. The method is not widely available in the clinic because of the limited accessibility of current flow-specific radiotracers. While the new flow tracer $^{18}$F-fluopiridaz \cite{Maddahi2020} does not have the accessibility problem, it may result in longer clinic visit times when combined with $^{18}$F-FDG for myocardial flow-metabolism imaging because of the long half-life of the $^{18}$F isotope in the two different tracers. 

The hypothesis of this study is that dynamic cardiac $^{18}$F-FDG PET imaging as a single tracer imaging can provide myocardial blood flow in addition to myocardial glucose metabolism by use of tracer kinetic modeling. The successful testing of this hypothesis may allow simultaneous imaging of myocardial blood flow and glucose metabolism only using $^{18}$F-FDG without the need for a second flow-specific tracer. Once validated, this single-tracer (i.e., FDG) multiparametric (i.e., flow and metabolism) imaging method has the potential to enable evaluation of myocardial viability and myocardial inflammation with reduced imaging time, cost and radiation exposure as compared to the traditional two-tracer methods in clinical practice today \cite{DiCarli2007, Maddahi2012}.  

The potential of $^{18}$F-FDG for blood flow imaging has been explored outside cardiac imaging. By use of tracer kinetic modeling \cite{Carson2005}, several studies have shown that the FDG blood-to-tissue delivery rate $K_1$ correlates with blood flow in tumors \cite{Tseng2004, Mullani2008, Bernstine2011, Cochet2012, Humbert2018}. For example, Tseng \emph{et al.}\cite{Tseng2004} demonstrated linear correlations between $K_1$ of 60-minute dynamic FDG-PET and $^{15}$O-water blood flow in breast tumors, with a linear correlation coefficient r=0.62 before neoadjuvant chemotherapy and r=0.81 after the therapy. Later, Mullani \emph{et al.} \cite{Mullani2008} reported that for various types of tumors in 16 patients, regional tumor FDG $K_1$ estimated from a 2-minute first-pass dynamic PET scan has a correlation r=0.86 with the blood flow measured by $^{15}$O-water PET. Correlation of FDG $K_1$ to blood flow has also been reported in organs such as liver and brain \cite{Winterdahl2011, Walberer2012}. In pigs, hepatic FDG $K_1$ derived from a 3-minute early-dynamic FDG-PET scan correlated with hepatic blood flow measured by transit time flow meters with a high correlation r=0.94  \cite{Winterdahl2011}. In a rat model of stroke, Walberer \emph{et al.} also reported that cerebral FDG $K_1$ estimated by one-hour dynamic PET data had a correlation of r=0.86 with $^{15}$O-water blood flow \cite{Walberer2012}. These studies support the potential use of $^{18}$F-FDG for estimating blood flow, though the \txtc{usability} of $^{18}$F-FDG $K_1$ likely depends on the specific tissue types; this is because FDG extraction fraction varies in different tissue {and is also dependent on blood flow}. 

Our work is specifically focused on \emph{myocardial} blood flow imaging using $^{18}$F-FDG. There is no prior study yet attempting to demonstrate the effectiveness of FDG flow in the myocardium. The importance of this work relies in the possible application of myocardial FDG flow to myocardial flow-metabolism mismatch evaluation for myocardial viability and inflammation and potentially more broadly, to rest-stress perfusion imaging for diagnosis of ischemic heart disease. Toward that end, our previous work specifically evaluated the practical identifiability of myocardial FDG $K_1$ quantification under different scan durations \cite{Zuo2020}. The results showed it is feasible to quantify FDG $K_1$ in the myocardium using appropriate kinetic modeling. 

The purpose of this paper is to directly compare myocardial FDG $K_1$ with MBF that is determined by a flow-specific tracer in human patients with heart disease. One challenge for using FDG to assess myocardial blood flow is the \txtc{potential correlation} of this glucose analog with blood glucose levels. As demonstrated in this paper, this may in turn compromise the performance of FDG $K_1$ for deriving MBF. To address this challenge, we also propose glucose normalization approaches to adjusting FDG $K_1$ that can reduce the effect of blood glucose levels. 

 \section{Methods}
 
 \subsection{Dynamic $^{18}$F-FDG PET and Dynamic $^{82}$Rb PET Data Acquisition}
 
Fourteen patients with ischemic heart disease (IHD) or suspected cardiac sarcoidosis were referred for PET myocardial viability or inflammation assessment by PET and enrolled into this study after giving informed consent. The study is approved by Institutional Review Board at the University of California, Davis. Each patient underwent a dynamic $^{82}$Rb-PET/CT scan {first} and {then} a dynamic FDG-PET/CT scan, both scans operated on a GE Discovery ST PET/CT scanner in two-dimensional mode. {The time between the Rb scan and FDG scan ranges from a few minutes to one hour, depending on the oral glucose loading procedure.}

{For dynamic Rb-PET imaging, patients received approximately 30 mCi $^{82}$Rb-chloride with a bolus injection. A low-dose transmission CT scan was performed at the beginning of PET scan to provide CT images for PET attenuation correction. The dynamic scan lasted for nine minutes. The acquired raw data were binned into 16 dynamic frames: 12 $\times$ 10 s, 2 $\times$ 30 s, 1 $\times$ 60 s, 1 $\times$ 300 s. Dynamic Rb-PET images were reconstructed using the standard ordered-subset expectation-maximization (OSEM) algorithm. All data corrections, including normalization, dead-time correction, attenuation correction, scatter correction, and random correction, were included in the reconstruction process. }

{For dynamic FDG-PET imaging, patients received approximately 20 mCi $^{18}$F-FDG with a bolus injection.  Data acquisition was commenced right after the FDG injection and lasted for 60 minutes. The acquired raw data were then binned into a total of 49 dynamic frames: 30 $\times$ 10 s, 10 $\times$ 60 s and 9 $\times$ 300 s.  Other processing was the same as for the Rb-PET scans. }

 \subsection{Kinetic Modeling of Cardiac $^{18}$F-FDG PET Data}

{An ellipsoidal region of interest (ROI) was manually placed in the left ventricle (LV) to extract an image-derived input function $C_\idif(t)$. An additional 17 ROIs were placed within the 17 segments of myocardium according to the AHA-17 standard \cite{Cerqueira2002}. These segment ROIs were combined into a global myocardial ROI and used to extract a global myocardial TAC $C_T (t)$ using ROI mean. Due to high noise of the dynamic data in individual segments, the analysis of this study was focused on global myocardial quantification.} Based on the analysis in our previous work \cite{Zuo2020}, a reversible two-tissue compartmental model \cite{Carson2005} was used to model the one-hour dynamic FDG-PET data:
\begin{eqnarray}
\frac{dC_1 (t)}{dt} &=& K_1 C_{p}(t)-(k_2+k_3 ) C_1(t)+ k_4 C_2(t),\\
\frac{dC_2 (t)}{dt}&=&k_3  C_1(t)- k_4 C_2(t),
\label{eq-2t}
\end{eqnarray}
where $C_p(t)$ is the FDG concentration in the plasma,  $C_1(t)$ is the concentration of free FDG and $C_2(t)$ denotes the activity concentration of metabolized tracer in the myocardium tissue space. $K_1$ is the tracer delivery rate from the blood space to the tissue space  with the unit \txtc{mL/min/cm$^3$\cite{Innis2007}};  $k_2$ (/min) is the rate constant of tracer exiting the tissue space; $k_3$ (/min) is phosphorylation rate;  $k_4$ (/min) is the dephosphorylation rate. {With $v_{b}$  denoting the fractional blood volume and $C_{wb}(t)$ denoting the activity in the whole blood},  the total activity that is measured by PET is described by
\begin{equation}
C_T(t) =(1-v_{b})[C_1(t) + C_2(t)]+v_{b}C_{wb}(t).
\label{eq-2tct}
\end{equation}
The unknown kinetic parameters ($v_{b}$,  $K_1$, $k_2$, $k_3$, $k_4$) were estimated using nonlinear least-square curve fitting. Their initial values were set to (0.1,0.1,0.1,0.1,0.001). The lower bounds were zero and the upper bounds were set to (1.0,2.0,2.0,1.0,0.1) \cite{Zuo2020}. 

\txtc{In theory the IDIF represents the activity in the whole blood, i.e., $C_\idif(t)=C_{wb}(t)$. The plasma input function $C_{p}(t)$ is related to $C_{wb}(t)$ following the model
\beq
C_{p}(t)=\pbr(t) \cdot C_{wb}(t),
\eeq
where $\pbr(t)$ denotes the plasma-to-blood ratio (PBR) function. In this work, we first used $\pbr(t)=1$ by assuming the difference between  $C_{p}(t)$ and $C_{wb}(t)$ is small \cite{Gambhir1989}.  Unless specified otherwise,  the main results of this paper were obtained with $\pbr(t)=1$.  In addition, we investigated the effect of PBR correction using a recent nonlinear model \cite{Naganawa2020},
\beq
\pbr(t)= 1/[0.97-0.06\exp(-0.085t)].
\label{eq-pbr2}
\eeq
}

\subsection{{Adjustment of FDG $K_1$ with Glucose Normalization}}

\txtc{We describe the relation between FDG $K_1$ and the blood glucose level $C_\glu$ using a linearized approximation,}
\beq
K_1\triangleq f(C_\glu) \approx f(C_{\glu,0})+\dot{f}(C_{\glu,0})\cdot\left(C_\glu-C_{\glu,0}\right),
\label{eq-K1exp}
\eeq
where $f(\cdot)$ is a nonlinear function of $C_\glu$ determined from the Renkin-Crone (RC) model and Michaelis-Menten model \cite{Carson2005}, see Eq. (\ref{eq-K1fun}) in the Appendix. $C_{\glu,0}$ denotes a reference blood glucose level and $\dot{f}(\cdot)$ is the first-order derivative of $f$. Based on \txtc{the derivations} in the Appendix I, we have
\beq
\dot{f}(C_{\glu,0})<0.
\eeq
This suggests FDG $K_1$ is inversely correlated with the blood glucose level,
\beq
K_1\approx -s\cdot C_\glu+int,
\label{eq-K1BG}
\eeq
where $s\triangleq -\dot{f}(C_{\glu,0})$ denotes the absolute slope ($s>0$) and $int$ is the intercept of the plot of FDG $K_1$ with respect to $C_\glu$. Practically, $s$ can be estimated using a correlation analysis between the measured FDG $K_1$ and $C_\glu$ data.

To reduce the dependency of FDG $K_1$ on $C_\glu$,  we consider $f(C_{\glu,0})$ as an adjusted FDG $K_1$. Based on  Eq. (\ref{eq-K1exp}), it can be calculated using the following form
\beq
K_1^{\adj} \triangleq f(C_{\glu,0})=K_1 + s\cdot\left(C_\glu-C_{\glu,0}\right),
\label{eq-K1adj}
\eeq
in which the blood glucose level of a patient is normalized and used to adjust the value of FDG $K_1$ linearly.
A similar linear correction was also used by Stout {\em et al.} \cite{Stout1998} for reducing the effects of normal physiological concentration of plasma large neutral amino acids on the $K_1$ of 6-[F-18]Fluoro-L-3,4-dihydroxyphenylalanine (FDOPA). As normal blood glucose levels range between 80-120 mg/dL, we use $C_{\glu,0}=100$ mg/dL in this work.

\txtc{Note that Eq. (\ref{eq-K1adj}) is an additive adjustment, in which $s$ is fixed and is independent of individual patients. Alternatively, a varying $s$, denoted as $s'$, may be used on patient basis,
\beq
s' = \frac{K_1}{C_{\glu,0}},
\eeq  
which results in a multiplicative adjustment,
\beq
K_1^{\adj} =\frac{C_{\glu}}{C_{\glu,0}}K_1.
\eeq
This approach assumes the product of $K_1$ and $C_\glu$ remains constant across different glucose levels. 
}

\subsection{Reference MBF by Dynamic Rb-PET}

Similar to the analysis of dynamic FDG-PET data described above, ROIs were drawn in the LV cavity and myocardium regions to extract regional TACs from the dynamic Rb-PET images. The myocardial TAC was modeled using a one-tissue compartmental model \cite{Lortie2007} with the following expression: 
{\begin{equation}
C_T(t) =(1-v_{b})C_1(t)+v_{b}C_{wb}(t),
\end{equation}
where $C_1(t)$ denotes the concentration of $^{82}$Rb in the tissue compartment with $K_1$ and $k_2$ now representing the rate constants of $^{82}$Rb transport between the plasma space and tissue space. This one-tissue model is equivalent to the two-tissue model in Eq. (\ref{eq-2t}) with $k_3=0, k_4=0$.} All the $^{82}$Rb kinetic parameters ($v_{b}$,  $K_1$, $k_2$) were estimated using nonlinear least-square curve fitting in a way similar to \cite{Zuo2020}.

The estimated Rb $K_1$ was then converted into myocardial blood flow $\mbf$ using the generalized RC model following Lortie's formula for $^{82}$Rb-PET data \cite{Lortie2007}:
{\beq
K_{1,\rb}= \mbf \cdot \left[1-a_\rb\cdot \exp\left(-\frac{b_\rb}{\mbf}\right)\right]
\label{eq-RbConvt}
\eeq
where $a_\rb = 0.77$ and $b_\rb = 0.63$ (mL/min/cm$^3$). MBF is derived from Rb $K_1$ through a look-up table that is pre-generated using the model.}

\subsection{{Extraction Fraction Correction (EFC) for Converting FDG $K_1$ to $ \mbf$}}

{Similar to the conversion Eq. (\ref{eq-RbConvt}) for $^{82}$Rb-chloride,  we can apply the generalized RC model to fit the FDG $K_1$ and MBF data under the assumption that capillary recruitment is involved at higher flow\cite{Yoshida1996} \txtc{(see Appendix II for further explanation)},
\begin{equation}
K_{1,\fdg}= \mbf \cdot\left[1-a_\fdg\cdot \exp\left(-\frac{b_\fdg}{\mbf}\right)\right],
\label{eq-rcfdg}
\end{equation}
where $a_\fdg$ and $b_\fdg$ are to be estimated from the paired data of FDG $K_1$ and Rb MBF using nonlinear least square fitting. The initial values for the two parameters $a_\fdg$ and $b_\fdg$ were both set to 0.1.  Note that if $a_\fdg$ is fixed at 1.0, then the model is equivalent to the classic RC model without capillary recruitment. }

The first-pass extraction fraction of FDG is defined by
\begin{equation}
E_\fdg\triangleq\frac{K_{1,\fdg}}{\mbf}.
\label{eq-exfdg1}
\end{equation}
Theoretically it relates to MBF following the generalized RC model,  i.e.,
\beq
E_\fdg=1-a_\fdg\cdot \exp\left(-\frac{b_\fdg}{\mbf}\right).
\label{eq-exfdg2}
\eeq

{The effect of flow-dependent FDG extraction fraction can be corrected using the inverse function of Eq. (\ref{eq-rcfdg}), leading to more quantitative FDG-derived MBF from FDG $K_1$. Similar to $^{82}$Rb-chloride, the nonlinear conversion  from FDG $K_1$ to MBF is performed through a look-up table using Eq. (\ref{eq-rcfdg}) with predetermined $a_\fdg$ and $b_\fdg$. The EFC was separately performed for the original FDG $K_1$ (i.e., without glucose normalization) and the adjusted FDG $K_1$ (i.e., with glucose normalization).}

\subsection{Statistical Analysis} 

We used the Pearson's correlation analysis and/or the Spearman correlation when appropriate to analyze the potential correlation between FDG $K_1$ and MBF and biological variables such as age (years), body mass index (BMI) (kg/m$^2$), and blood glucose (BG) level (mg/dL). A p value $\le$0.05 was considered as statistically significant. All the analyses were done using MATLAB (MathWorks, MA). \txtc{The Bland-Altman plot \cite{Bland1986} was used to quantitatively compare FDG MBF with Rb MBF. When appropriate, the estimated standard error (SE) of a parameter estimate $x$ was also reported in the format of $x\pm$SE.}

\section{Results}

\subsection{Patient Characteristics}

Patient characteristics are provided in table \ref{tbl:patients}. Among the fourteen patients enrolled in the study, ten were diagnosed as ischemic heart disease and four were diagnosed with or suspected of cardiac sarcoidosis prior to the scans.  All of the patients completed the dynamic $^{82}$Rb-PET scan.  Twelve patients had a dynamic FDG-PET scan of 50-60 minutes. Two other patients had a dynamic FDG-PET scan of only 30-40 minutes due to discomfort,  and hence these two subjects were not included in this study. 

Other characteristics of the patients, including age, sex, diabetic status, BMI, BG level (before PET imaging), and dynamic FDG-PET scan duration are also reported in table \ref{tbl:patients}. Unavailable data is marked with `/' in the table.
 
 \subsection{Myocardial TAC Fitting and Kinetics}

\begin{figure*}[t]
 \centering
\subfigure[]{
 \includegraphics[trim=0cm 0.0cm 1.3cm 0cm, height=5.5cm]{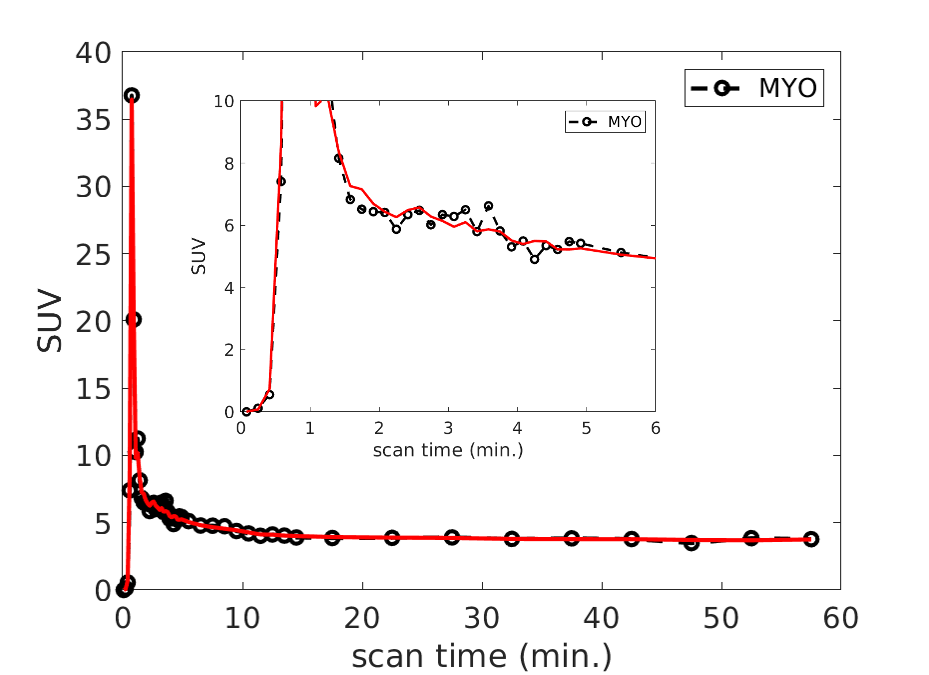}}
 \subfigure[]{
\includegraphics[trim=0cm 0.0cm 1.3cm 0cm, height=5.5cm]{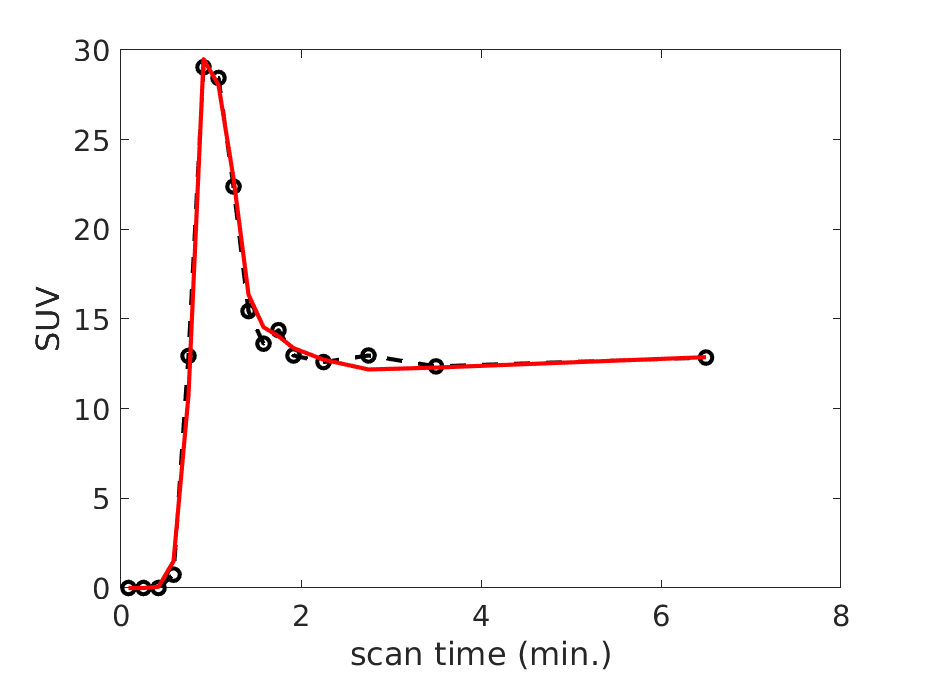}}
  \caption{Example of myocardial TAC fitting of the FDG \txtc{and Rb data. }(a) FDG; (b) Rb.}
 \label{fig:TAC_fitting}
 \end{figure*}

\txtc{Fig. \ref{fig:TAC_fitting} (a) and (b) show an example of myocardial TAC fitting for the FDG data and Rb data, respectively. The fits demonstrated a good match between the measured time points and predicted TAC by the model in each case. }
The estimated $^{18}$F-FDG kinetic parameters and $^{82}$Rb-chloride MBF values of all the patients are summarized in table \ref{tbl:eg_Ks_FDG}. 
For the FDG protocol, the results of two patients were not available due to an incomplete dynamic FDG scan.  The average SE values across the twelve patients are reported in table \ref{tbl:eg_biasSD} for each of the FDG kinetic parameters. The result indicates that FDG $K_1$ was able to be estimated with a low SE ($<14\%$).

 \begin{table}
\caption{Characteristics of the patients enrolled in the study. \\} 
\centering
\footnotesize
\begin{tabular}{lcccccccc}
\hline 
Patient &IHD& Age (years)  & Sex &Diabetic & BMI & BG (mg/dL) & FDG Scan Time (min)\\
\hline
1 & Y                                  & 58   &M    &Y    &38.6    &127   & 60\\
2 & Y                                 &73    &M    &N    &24.4    &113    & 50\\
3 & Y                                  & 61   &M    &N    &33.9    &88      &60\\
4 & N    & 71   &F     &N    &22.4    &116    &60\\ 
5 & Y                                  & 55   &F     &N    &24.5    &118    &40\\
6 & N                       & 57   &M    &N    &27.4    &/         &60\\
7 &  Y                                 &  63  &M    &N    &28.8    &105   &60\\
8 &  Y                                 & 59   &M   &Y   &28.3   &135  &60\\
9 &  Y                                 & 65   &M   &Y    &27.2   &130   &50\\
10 & N                     & 81   &M   &N   &25.8   &85      &60\\
11 & Y                               & 74   &M  &N   &28.2    &84      &60\\
12 & Y                                & 83   &M  &N   &33.6    &/         &60\\
13 & Y                                & 59   &M  &Y    &35.9    &107    &60\\
14 & N                     & 69   &F    &N   &31.4    &82     &30\\
\hline
\end{tabular}
\label{tbl:patients}
\end{table}

\begin{table}
\caption{{Estimates of FDG kinetics and $^{82}$Rb-chloride MBF\\}}
\centering
\footnotesize
\begin{tabular}{lccccccccccc}
\hline
& \multicolumn{5}{c}{$^{18}$F-FDG} && \multicolumn{2}{c}{$^{82}$Rb-chloride} && \multicolumn{2}{c}{Adjusted FDG $K_1$} \\
\cline{2-6}\cline{8-9}\cline{11-12}
No. & $v_{b}$ & $K_1$ & $k_2$ & $k_3$ & $k_4$ && $v_b$ & MBF &&  \txtc{additive} & \txtc{multiplicative}\\\hline
1   &    0.603    & 0.138     & 0.399     & 0.021     & 0.027 && 0.446 & 0.440 && 0.293 & 0.175\\
2  &     0.249   & 0.440     & 1.068     & 0.129    &  0.017  &&  0.374 & 0.651 &&  0.515& 0.498\\
3  &     0.380    & 0.675     & 2.000     & 0.176     & 0.022 &&  0.389 & 0.904 && 0.607& 0.594\\
4  &     0.279     &    0.681    &  2.000     & 0.061    &  0.027 &&    0.304 & 1.842 &&  0.773& 0.790\\
5  &       /             &         /           &       /           &       /          & /  &&  0.366 &    0.674 && / &  /\\
6  &     0.368     &      0.593     & 2.000     & 0.041    & 0.081 &&  0.411 &    0.812 &&     /& /\\
7  &     0.494     &     0.279    &  1.059     & 0.091     & 0.054 && 0.377 &    0.432 && 0.308& 0.293\\
8  &     0.371       &   0.284     & 0.690     & 0.107     & 0.043 &&0.362 &     0.642 &&  0.485& 0.384\\
9  &     0.461        &   0.366    &  0.795     & 0.056    &  0.017 && 0.336 &   0.700 &&     0.538& 0.476\\
10  &     0.328       &    0.667    &  1.976     & 0.222     & 0.031 &&0.359 &     0.953 && 0.581 & 0.567\\
11 &     0.260     &      0.413     & 1.266     & 0.072     & 0.026 && 0.311 &   0.617 && 0.321& 0.347\\
12 &    0.391     &      0.584    & 1.539     & 0.061    &  0.032 &&0.294 &    0.633 && / & /\\
13 &     0.591     &    0.302    &  0.616     & 0.046    &  0.026 &&0.502 &  0.541 && 0.342 & 0.323\\
14  &       /             &         /           &       /           &       /          & /  && 0.428 &      1.004 && /& /\\
\hline
\end{tabular}
\label{tbl:eg_Ks_FDG}
\end{table}

\begin{table}
\caption{\txtc{Average standard error (SE) of the estimates of FDG kinetics\\} }
\centering
\footnotesize
\begin{tabular}{lccccc}
\hline
FDG &$v_{b}$ & $K_1$ & $k_2$ & $k_3$ & $k_4$ \\\hline
SE  (\%)  & 3.5  & 13.8& 15.4 & 18.4& 19.7\\
\hline
\end{tabular}
\label{tbl:eg_biasSD}
\end{table}

\begin{table}
\caption{Pearson's r and p of FDG $K_1$ and MBF with patients' age, BMI, and blood glucose (BG) concentration.\\} 
\centering
\footnotesize
\begin{tabular}{l | llcccc}
\hline\hline 
Biological Variables &&Age&BMI &BG\\
\hline\hline
\multirow{2}*{FDG $K_1$} &r & 0.495   & -0.443   &-0.562   \\
&p & 0.102   & 0.150     &0.091  \\\hline
\multirow{2}*{$^{82}$Rb MBF}& r &0.237   & -0.565  &-0.076 \\
&p & 0.458   & 0.055     &0.835 \\
\hline\hline
\end{tabular}
\label{tbl:r_p_MBF_others}
\end{table}

\subsection{{Correlation of FDG $K_1$ with Blood Glucose and Other Biological Variables}}

Table \ref{tbl:r_p_MBF_others} summarizes the Pearson's r and p values between FDG $K_1$ and the biological variables including age,  BMI, and BG level. The results of MBF were also included in the table for comparison. The estimated fractional blood volume $v_b$ of FDG was not exactly equal to the $v_b$ of Rb as shown in table \ref{tbl:eg_Ks_FDG} but the two parameters were correlated with each other (r=0.696, p=0.012).

\begin{figure*}[t]
 \centering
 \subfigure[]{\includegraphics[trim=0.1cm 0cm 1.2cm 0cm, clip,height=4.0cm]{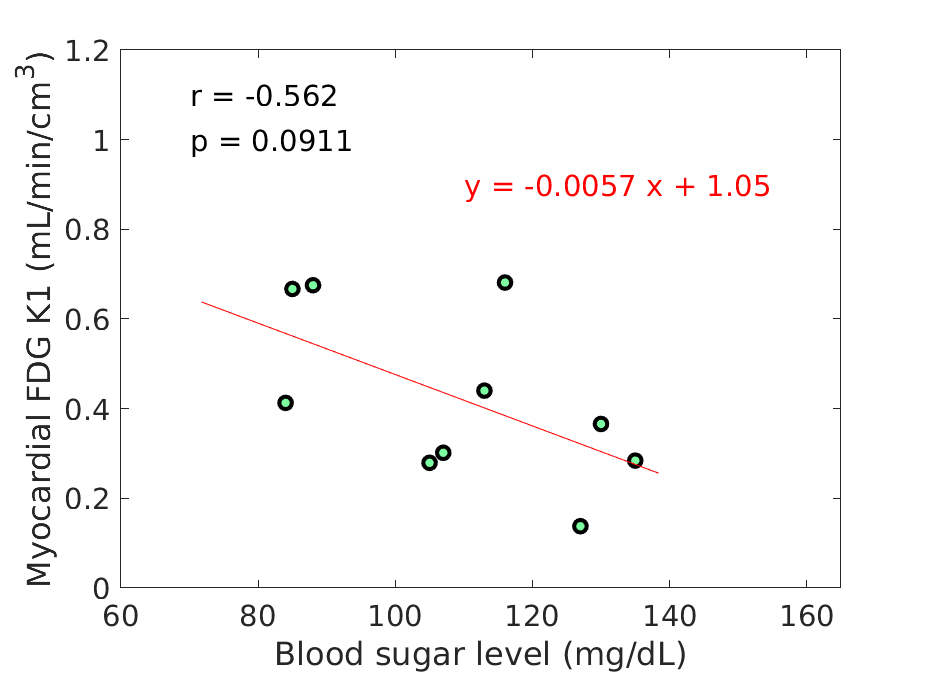}}
 \subfigure[]{\includegraphics[trim=0.1cm 0cm 1.4cm 0cm, clip,height=4.0cm]{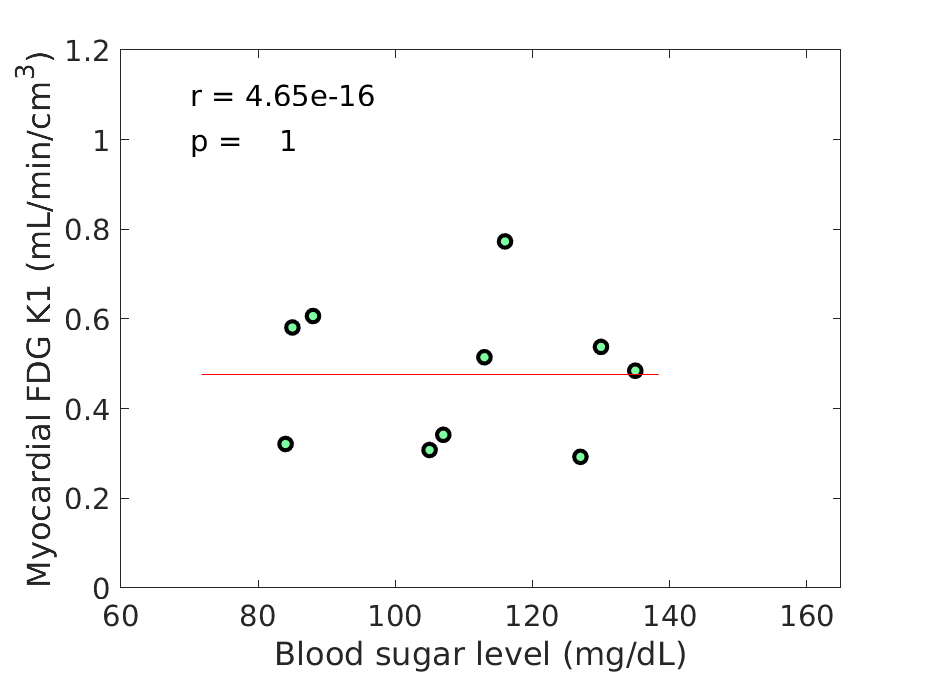}}
  \subfigure[]{ \includegraphics[trim=0.1cm 0cm 1.4cm 0cm, clip,height=4.0cm]{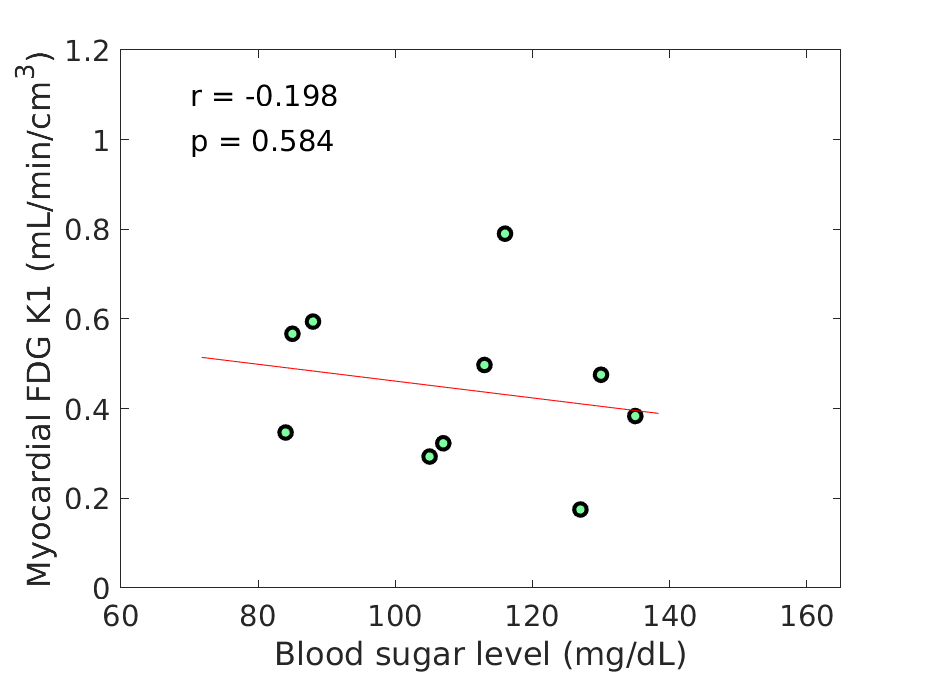}}

  \caption{Correlation between FDG $K_1$ and blood glucose level before and after FDG $K_1$ is adjusted for blood glucose levels. \txtc{(a) no adjustment, (b) by the additive adjustment, and (c) by the multiplicative adjustment.}}
 \label{fig:K1BG}
 \end{figure*}

FDG $K_1$ did not correlate with age and BMI. MBF tended to inversely correlate with BMI (r=-0.565, p=0.055). FDG $K_1$ tended to inversely correlate with BG (r=-0.562, p=0.091) {as also shown in Fig. \ref{fig:K1BG}(a), while a similar trend was not observed for MBF (r=-0.076, p=0.835).  The negative correlation between FDG $K_1$ and BG is \txtc{in line with} the derivation in Eq. (\ref{eq-K1BG}).}

The estimated slope $s$ between FDG $K_1$ and BG was 0.0057\txtc{$\pm$0.0030}, with which the original FDG $K_1$ was then adjusted for blood glucose levels using the additive adjustment Eq. (\ref{eq-K1adj}) with a reference $C_{\glu,0}=100$ mg/dL. The FDG $K_1$ after the adjustment is included in table \ref{tbl:eg_Ks_FDG}. \txtc{The multiplicative adjustment was also implemented and the result is included in table \ref{tbl:eg_Ks_FDG} as well.} Note that the adjustments were not applied if a BG value was unavailable. \txtc{Therefore the two patients (\#6 and \#12) were not included in the subsequent analysis.} The linear dependency of FDG $K_1$ on blood glucose diminished with glucose normalization. \txtc{Fig. \ref{fig:K1BG}(b-c) show that the negative correlation was no longer existing (by the additive adjustment) or reduced (by the multiplicative adjustment) between the adjusted FDG $K_1$ and blood glucose levels. These three sets of FDG $K_1$ were all used in the subsequent analysis. }

 \begin{figure*}[t]
 \centering
 \subfigure[]{
 \includegraphics[trim=0.2cm 0cm 1.4cm 0.5cm, clip,height=3.8cm]{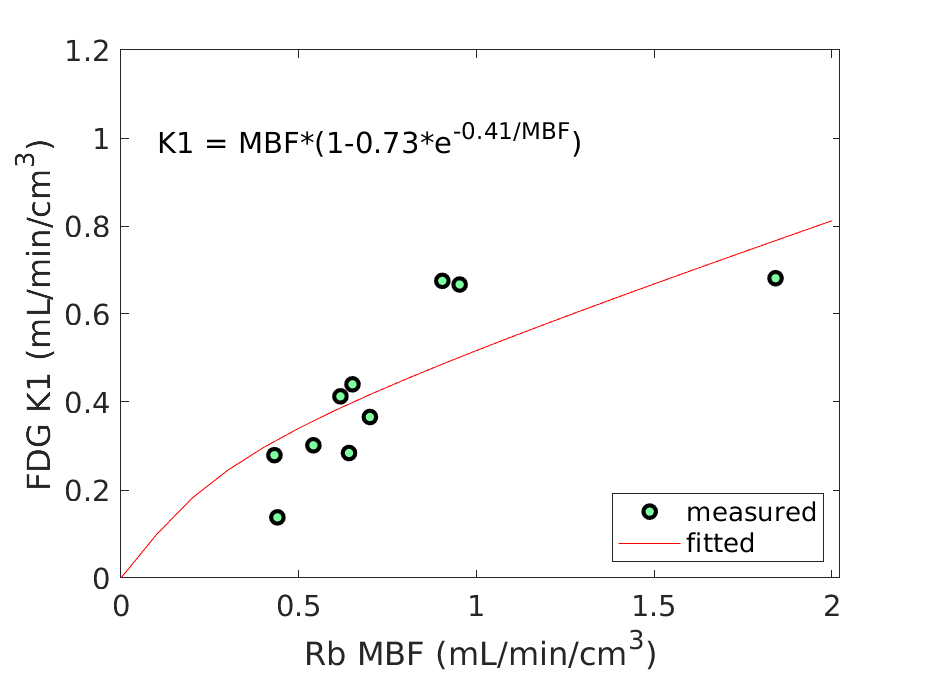}}
 \subfigure[]{
 \includegraphics[trim=0.2cm 0cm 1.4cm 0.5cm, clip,height=3.8cm]{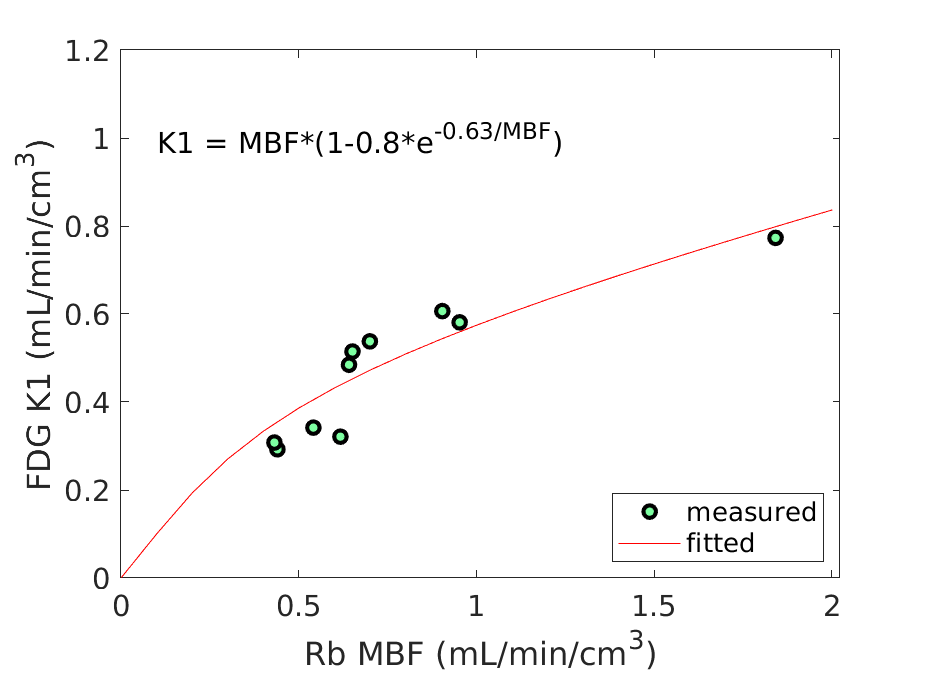}}
\subfigure[]{
 \includegraphics[trim=0.2cm 0cm 1.4cm 0.5cm, clip,height=3.8cm]{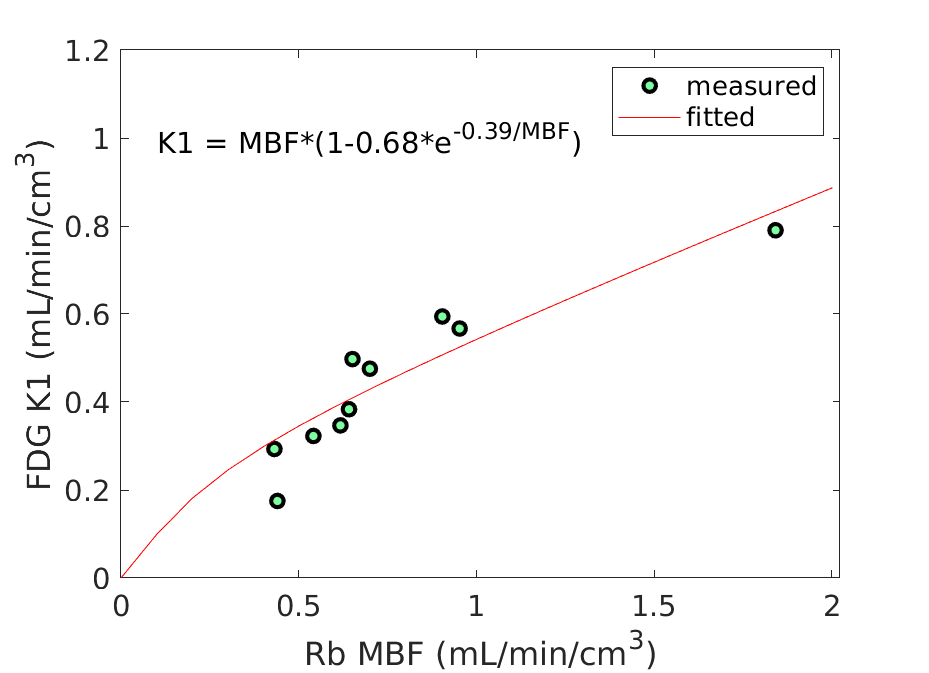}}
 \caption{Nonlinear association between FDG $K_1$ and MBF before and after FDG $K_1$ is adjusted for blood glucose levels. \txtc{(a) no adjustment, (b) by the additive adjustment, and (c) by the multiplicative adjustment.}}
 \label{fig:K1MBF}
 \end{figure*}

  \begin{figure*}[t]
 \centering
 \subfigure[]{ \includegraphics[trim=1.8cm 0cm 2.5cm 0cm, clip,height=5.1cm]{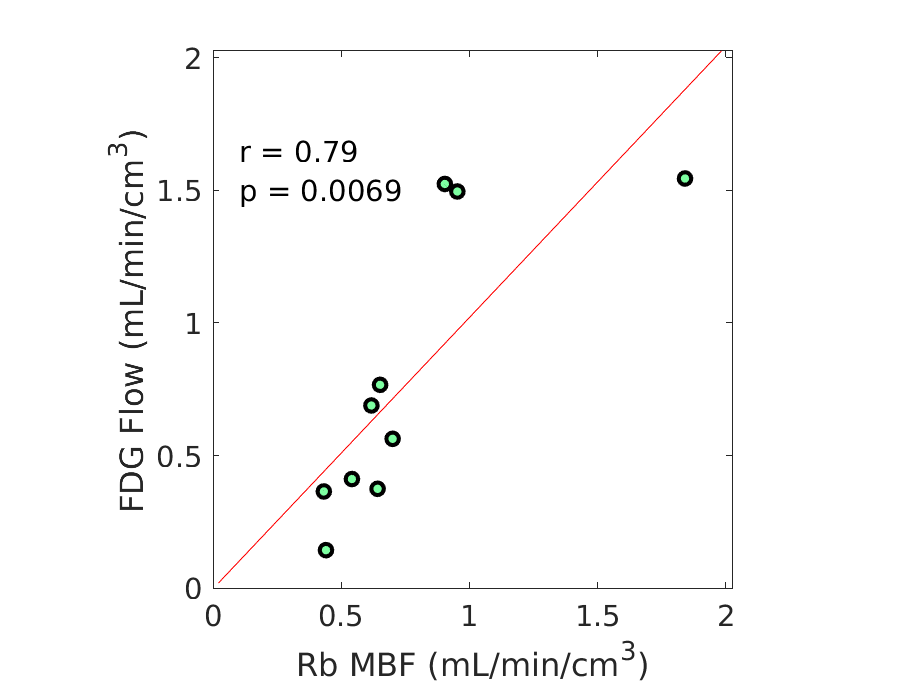}}
 \subfigure[]{ \includegraphics[trim=1.8cm 0cm 2.5cm 0cm, clip,height=5.1cm]{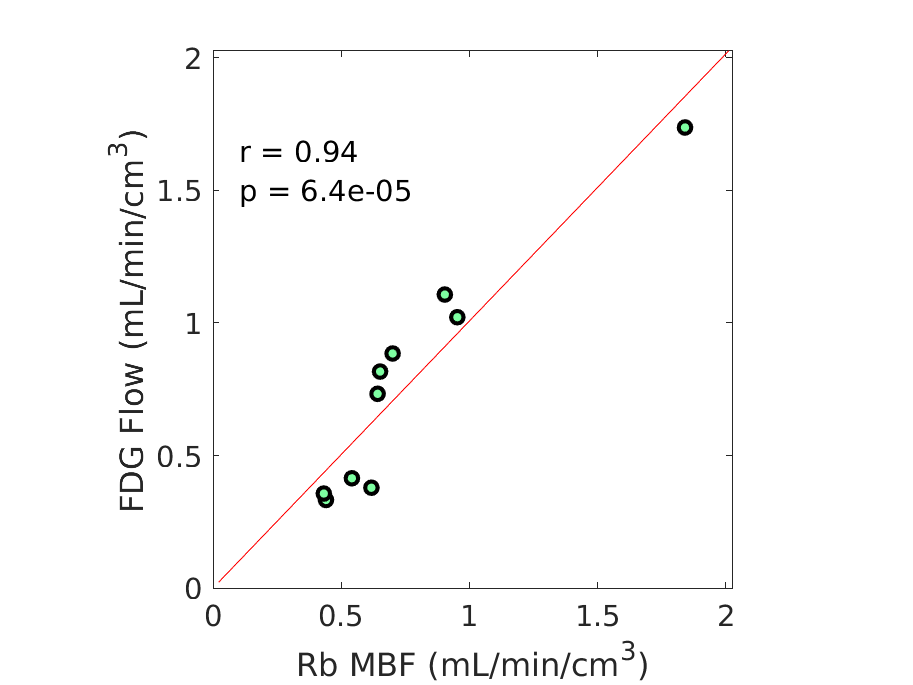}}
  \subfigure[]{ \includegraphics[trim=1.8cm 0cm 2.5cm 0cm, clip,height=5.1cm]{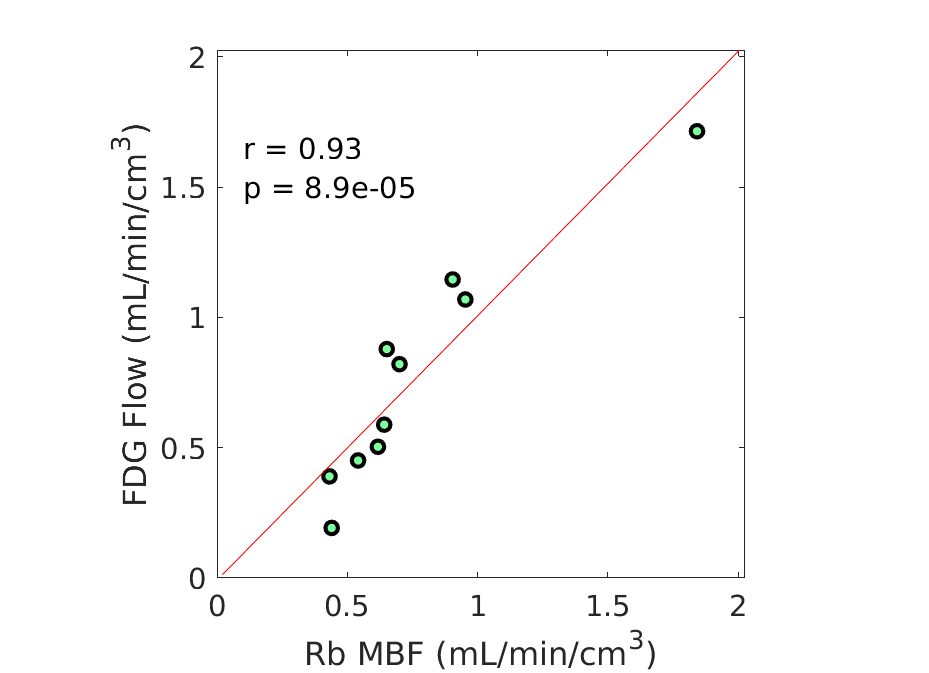}}
  \caption{Correlation between FDG MBF and Rb MBF before and after FDG $K_1$ is adjusted for blood glucose levels. \txtc{(a) no adjustment, (b) by the additive adjustment, and (c) by the multiplicative adjustment.}}
 \label{fig:FDGMBF}
 \end{figure*}

 \subsection{{Nonlinear Conversion from FDG $K_1$ to MBF}}
 
Fig. \ref{fig:K1MBF} shows the nonlinear association of FDG $K_1$ with MBF before and after the adjustment of FDG $K_1$ for blood glucose. \txtc{The Spearman correlation was 0.89 (p=0.0014) before the adjustment and became 0.96 (p$<$0.0001) after the additive or multiplicative adjustment.  For the three cases, the data were fitted using the generalized RC model Eq. (\ref{eq-rcfdg}).  The estimated model parameters were $a_\fdg=0.73\pm0.15$ and $b_\fdg=0.41\pm 0.21$ for the approach without adjustment, $a_\fdg=0.80\pm0.10$ and $b_\fdg=0.63\pm0.11$ for the approach with the additive adjustment, and $a_\fdg=0.68\pm0.09$ and $b_\fdg=0.39\pm0.14$ for the approach with the multiplicative adjustment. }

\begin{figure*}[t]
 \centering
 \subfigure[]{ \includegraphics[trim=0cm 0cm 1.2cm 0cm, clip,width=5.0cm]{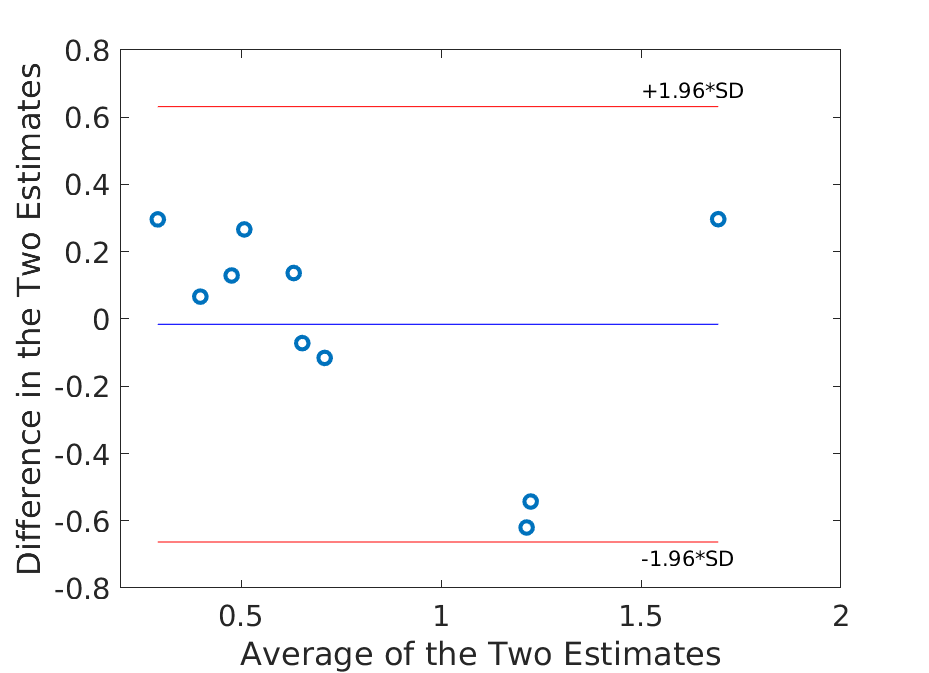}}
 \subfigure[]{ \includegraphics[trim=0cm 0cm 1.2cm 0cm, clip,width=5.0cm]{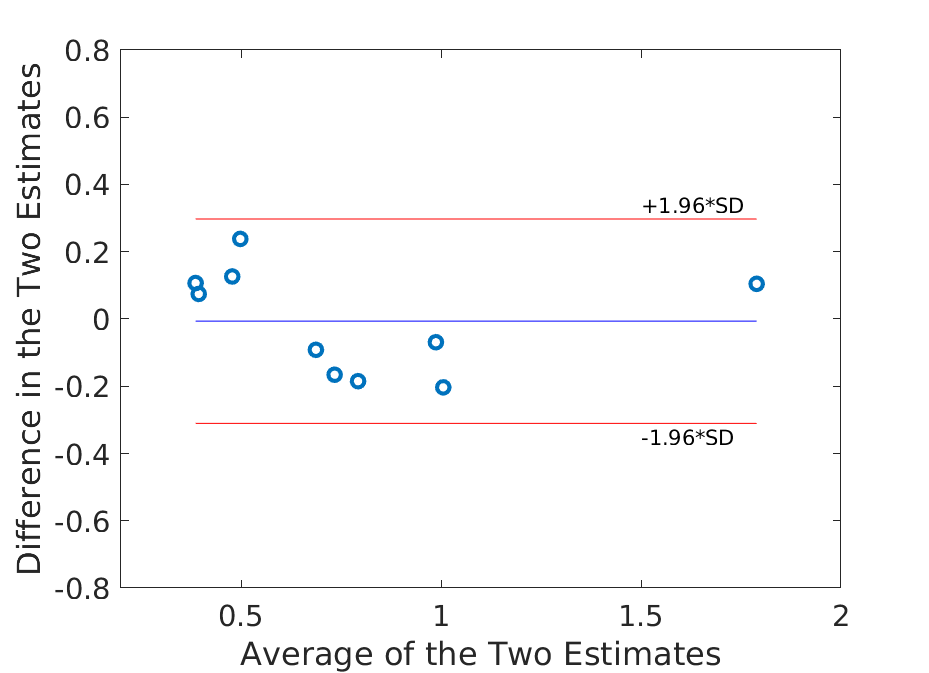}}
  \subfigure[]{ \includegraphics[trim=0cm 0cm 1.2cm 0cm, clip,width=5.0cm]{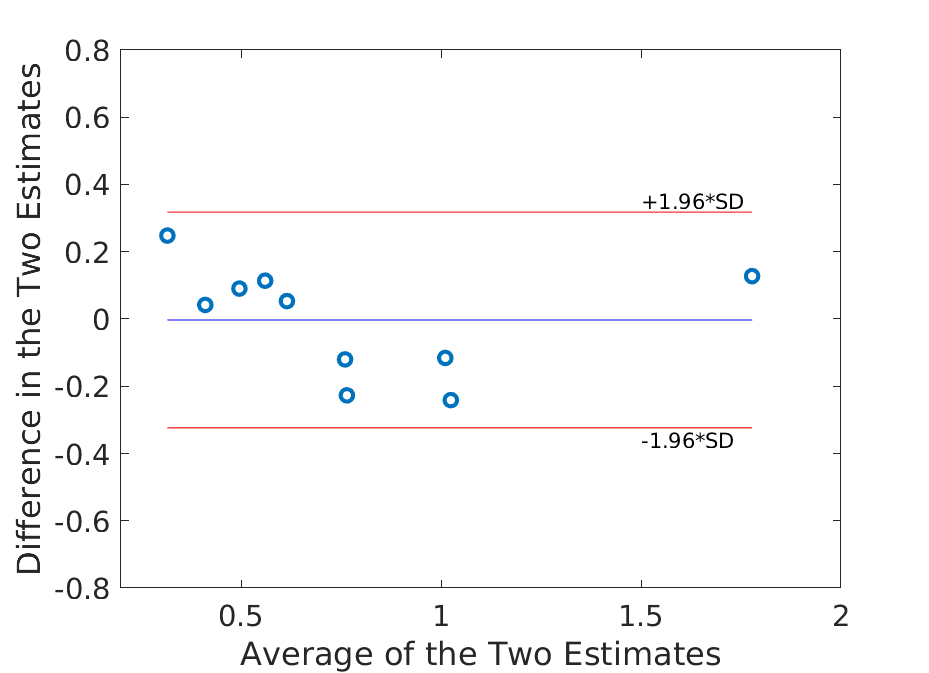}}
  \caption{ \txtc{Bland-Altman plots of FDG MBF and Rb MBF (unit: mL/min/cm$^3$) before and after glucose normalization. (a) no adjustment, (b) by the additive adjustment, and (c) by the multiplicative adjustment.}}
 \label{fig:BAPlot}
 \end{figure*}

Using the fitted model for extraction fraction correction, FDG $K_1$ was converted to MBF in each case based on Eq.  (\ref{eq-rcfdg}). Fig. \ref{fig:FDGMBF} shows the results of linear Pearson correlation between FDG-derived MBF and Rb MBF.  The three correlations were statistically significant (p$<$0.01).  \txtc{The correlation coefficient was moderate (r=0.79) without adjusting FDG $K_1$ for blood glucose, and became improved (r$>$0.9) after the additive or multiplicative adjustment. Fig. \ref{fig:BAPlot} furthers shows the Bland-Altman plots of the blood flow estimates. The differences between FDG MBF and Rb MBF were large without a glucose adjustment and became smaller with the adjustments.}

Table \ref{tbl:det_mean} further compares the statistics (mean and standard deviation) of Rb MBF and FDG-derived MBF in the ischemic heart disease (IHD) group (9 patients) and non-IHD group (3 patients).  The mean and standard deviation became closer to that of Rb MBF after the additive adjustment of FDG $K_1$ for blood glucose in both groups.

\begin{table}[t]
 \caption{{Mean and standard deviation of Rb MBF and FDG MBF in IHD and non-IHD patients. Unit: mL/min/cm$^3$\\} }
\centering
\footnotesize
\begin{tabular}{l|cc}
\hline\hline 
&IHD & non-IHD\\\hline
$^{82}$Rb MBF  & 0.618$\pm$0.143 & 1.202$\pm$0.558 \\
FDG MBF without $K_1$ adjustment & 0.609$\pm$0.433 & 1.373$\pm$0.190\\
FDG MBF with $K_1$ adjustment & 0.624$\pm$0.280 & 1.222$\pm$0.448\\
\hline\hline
\end{tabular}
\label{tbl:det_mean}
\end{table}

\begin{figure*}[t]
 \centering
   \subfigure[]{\includegraphics[trim=0.2cm 0cm 1cm 0.0cm, clip,height=4.1cm]{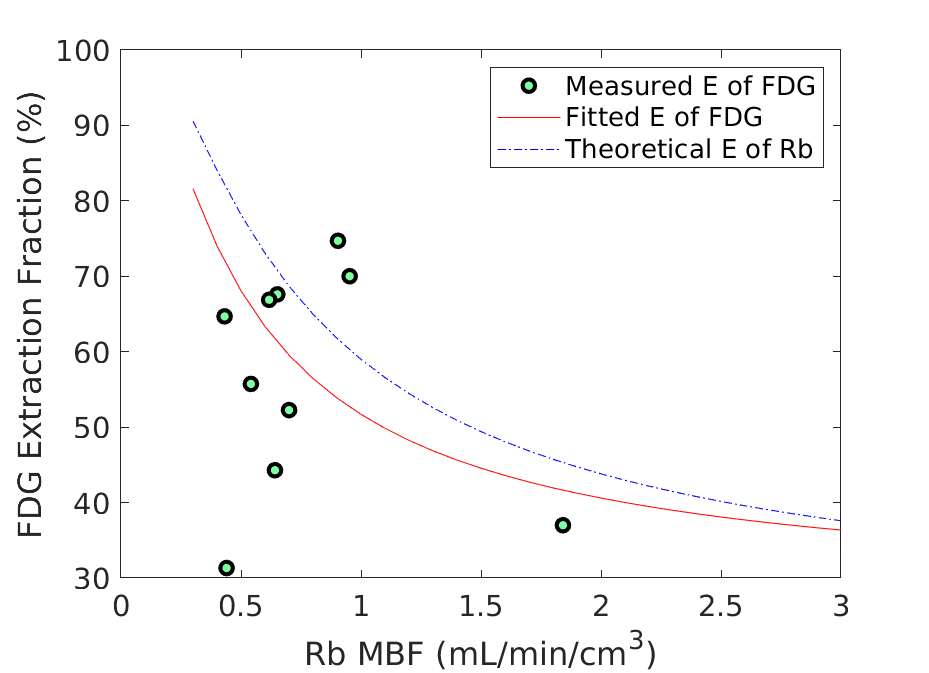}}
  \subfigure[]{\includegraphics[trim=0.2cm 0cm 1cm 0.0cm, clip,height=4.1cm]{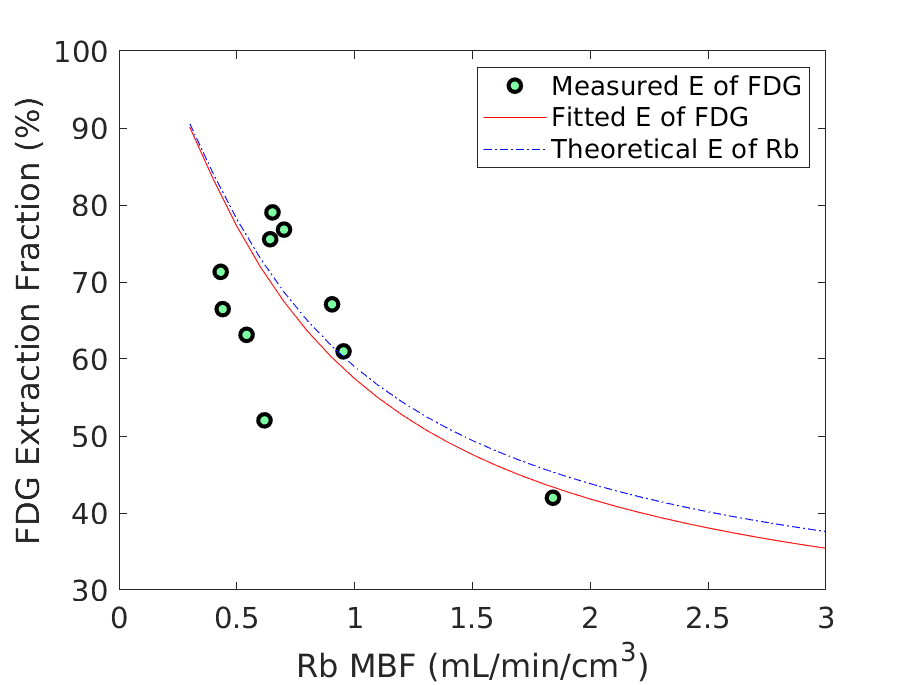}}
    \subfigure[]{\includegraphics[trim=0.2cm 0cm 1cm 0.0cm, clip,height=4.1cm]{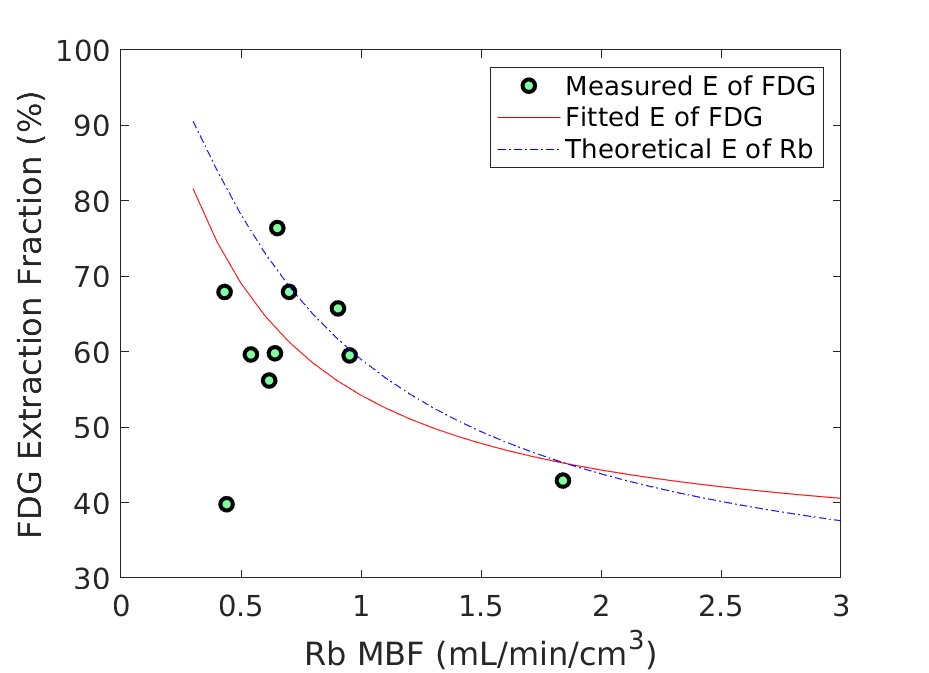}}
  \caption{Extraction fraction of FDG in the myocardium as compared to Rb MBF before and after glucose normalization. (a) no adjustment, (b) by the additive adjustment, and (c) by the multiplicative adjustment. }
 \label{fig:FDGext}
 \end{figure*}

\subsection{Myocardial Extraction Fraction of FDG}

Fig. \ref{fig:FDGext} shows the extraction fraction of FDG in the myocardium as a function of MBF using the FDG $K_1$ estimates with and without adjustment for blood glucose levels.  Both the measured $E_\fdg$ using Eq. (\ref{eq-exfdg1}) and the calculated model Eq. (\ref{eq-exfdg2}) are shown.  The theoretical extraction fraction of Rb-chloride in the myocardium \cite{Lortie2007} is also included for comparison.  The results show that the relation of the extraction fraction of FDG with respect to MBF was close to that of Rb-chloride in a range of MBF between 0-2 mL/min/cm$^3$. \txtc{It seems that the additive adjustment approach led to a slightly better fit to the generalized RC model than the multiplicative adjustment did. Overall the FDG extraction fraction was about 60\% relative to Rb MBF at 1.0 mL/min/cm$^3$, while Marshall {\em et al} \cite{Marshall1998} reported an extraction fraction of 70\% in rabbit heart.}

 \begin{figure*}[t]
 \centering
    \subfigure[]{\includegraphics[trim=0cm 0cm 1cm 0.0cm, clip,height=4.5cm]{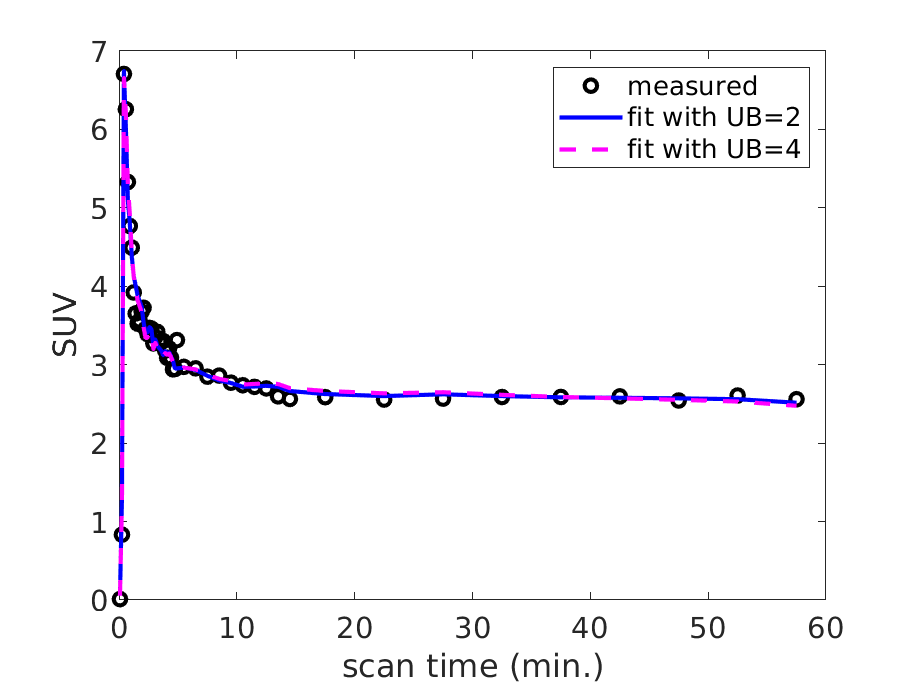}}
\subfigure[]{\includegraphics[trim=0cm 0cm 1cm 0.0cm, clip,height=4.5cm]{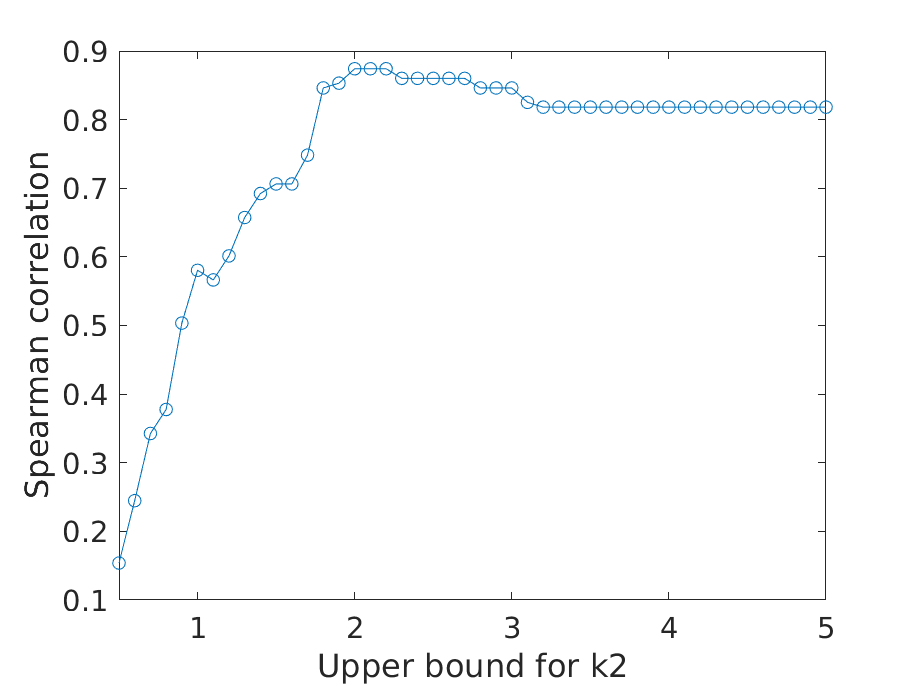}}
  \caption{\txtc{Effect of the FDG $k_2$ upper bound (UB). (a) myocardial TAC fitting examples with two different UB values in a patient; (b) effect of the UB on the Spearman correlation between FDG $K_1$ and Rb MBF in all patients.}}
 \label{fig:k2UB}
 \end{figure*}

 \subsection{\txtc{Effect of the upper bound for $k_2$}}

\txtc{As shown in table \ref{tbl:eg_Ks_FDG}, some of the FDG $k_2$ estimates hit the upper bound (UB) 2.0. With a higher UB, the $k_2$ estimates can go higher, which in turn increases the $K_1$ value due to the coupling between $K_1$ and $k_2$. Fig. \ref{fig:k2UB}(a) shows the examples with two UB values (2.0 and 4.0) for fitting a myocardial TAC. Both options fitted the TAC reasonably well. UB=4.0 provided a lower fitting error (mainly in the early time), which however may be because of over-fitting of the noise. Fig. \ref{fig:k2UB}(b) further shows the effect of the $k_2$ UB on the Spearman correlation of FDG $K_1$ with Rb MBF in all patients. The correlation reached the peak at UB=2.0 and became lower at higher UB values. The use of the upper bound can be explained as adding a regularization in TAC fitting, which has the role of preventing over-fitting and stabilizing the kinetic estimation. }
 
\subsection{\txtc{Effect of the plasma-to-blood ratio correction}}

\begin{figure*}[t]
 \centering
   \subfigure[]{\includegraphics[trim=0.8cm 0cm 2cm 0.0cm, clip,height=4.2cm]{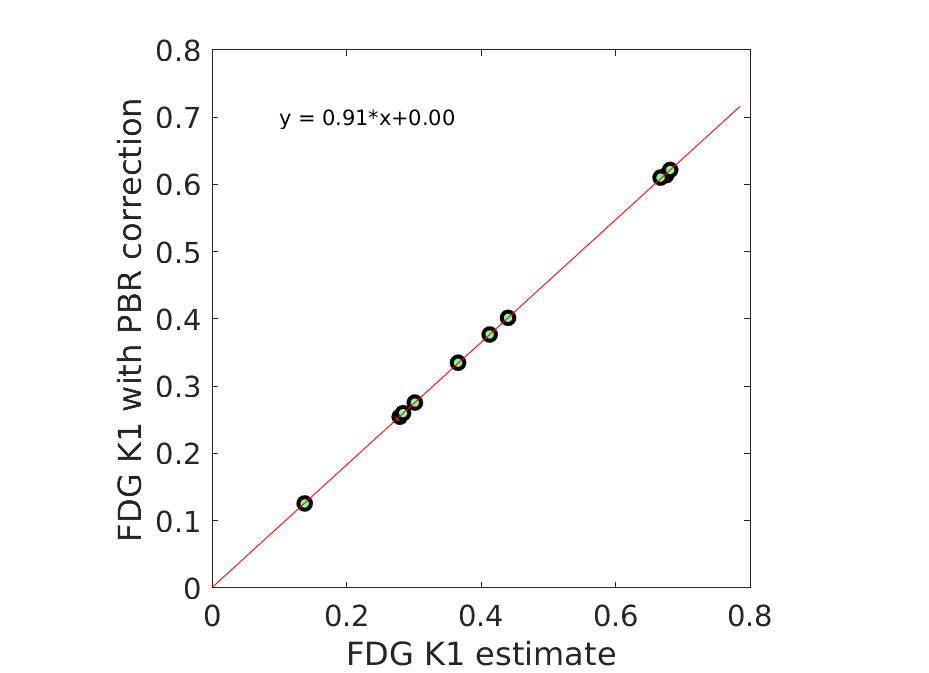}}
  \subfigure[]{\includegraphics[trim=0.2cm 0cm 1.3cm 0.0cm, clip,height=4.2cm]{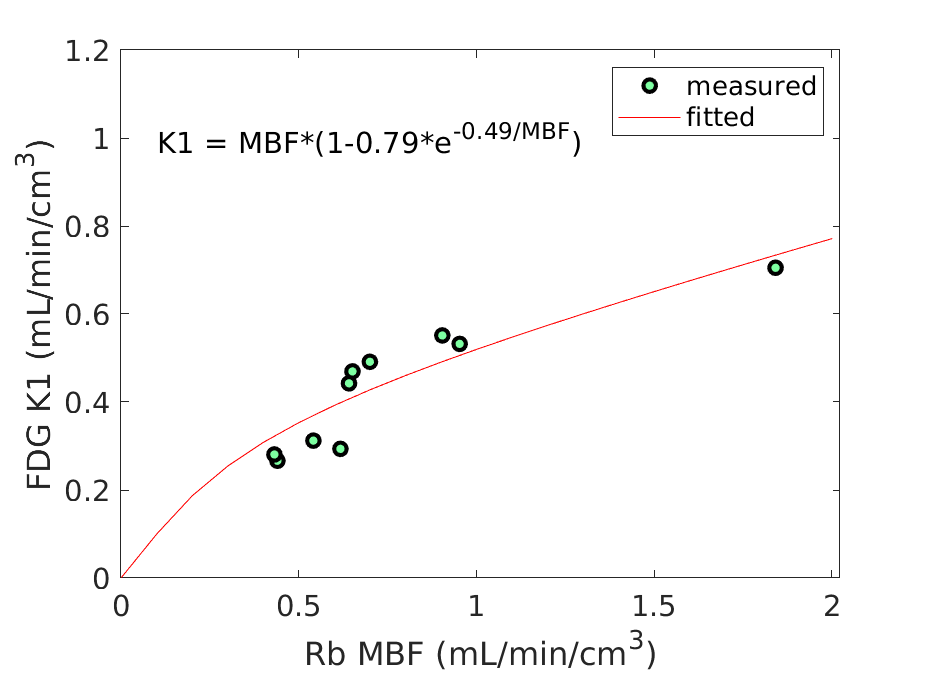}}
    \subfigure[]{\includegraphics[trim=0.0cm 0cm 1.3cm 0.0cm, clip,height=4.2cm]{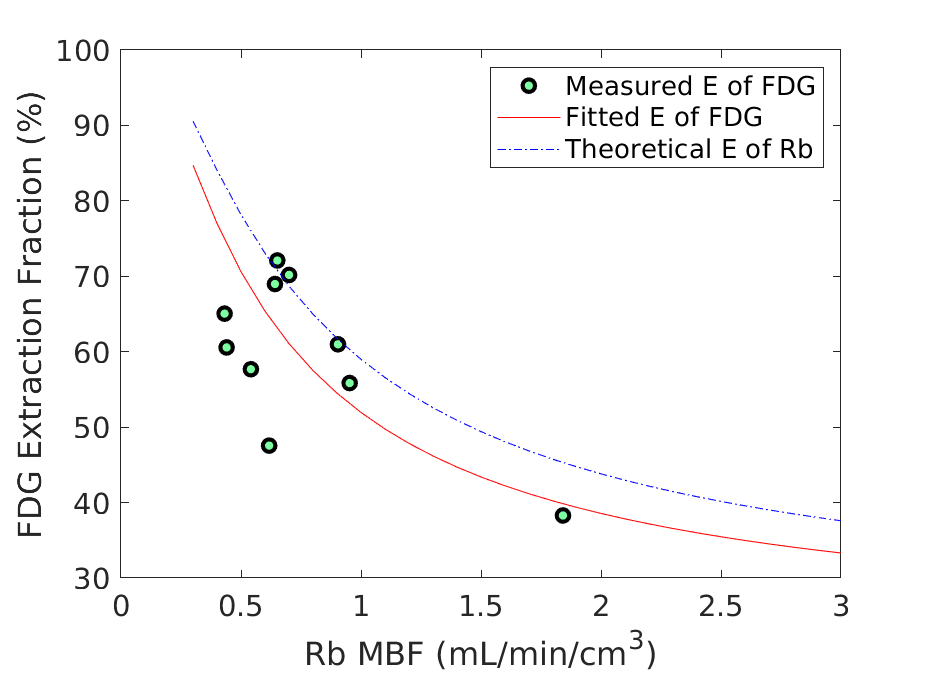}}
  \caption{\txtc{Effect of the plasma-to-blood ratio correction on FDG $K_1$. (a) the scaling effect between the FDG $K_1$ estimates with and without PBR correction; (b) RC model fit of the PBR-corrected FDG $K_1$ with respect to Rb MBF; (c) FDG extraction fraction calculated using the PBR-corrected FDG $K_1$.}}
 \label{fig:PBR}
 \end{figure*}

\txtc{All the results reported above were obtained with $\pbr(t)=1$, i.e. no PBR correction was used. The effect of using the PBR correction in Eq. (\ref{eq-pbr2}) is shown in Fig \ref{fig:PBR}(a). Here the FDG $K_1$ was obtained with the additive adjustment. The FDG $K_1$ with the PBR correction was related to the PBR-uncorrected $K_1$ by a scaling factor $0.911\pm0.002$. The result is consistent with the past studies that reported an average PBR of approximately 1.1 (see the references in \cite{Naganawa2020}), which in turn should result in a global scaling of 0.91 in the $K_1$ estimates. Fig \ref{fig:PBR}(b) shows the fit of the PBR-corrected FDG $K_1$ with respect to Rb MBF using the generalized RC model. The estimated FDG extraction fraction shown in Fig \ref{fig:PBR}(c) was lower when comparing to Fig \ref{fig:FDGext}(b). However, the resulting change (not shown) on the final FDG-derived MBF is negligible because the scaling factor was compensated in the generalized RC model.}

\txtc{Note that the lower FDG extraction fraction after the PBR correction may be not reflecting the ground truth because ideally the Rb data should also be corrected for PBR. However, it is practically difficult provided that there is no published model to use. }

\section{Discussion}

In this paper, we investigated the feasibility of $^{18}$F-FDG for assessing MBF through a pilot clinical study. The FDG delivery rate $K_1$ demonstrated an inverse correlation with blood glucose levels according to both the \txtc{analytical} investigation in Eq. (\ref{eq-K1BG}) and patient data shown in Fig. \ref{fig:K1BG}(a).  We therefore \txtc{studied two glucose normalization approaches to adjusting FDG $K_1$ for removing the dependence of FDG $K_1$ on blood glucose (Fig. \ref{fig:K1BG}).}  FDG $K_1$ was further compared to the Rb reference MBF for analyzing the relation between them using the generalized RC model (Fig. \ref{fig:K1MBF}). The resulting EFC model was then used to convert FDG $K_1$ to MBF. The results showed that the FDG-derived MBF correlated well with Rb MBF and glucose normalization for FDG $K_1$  was important (Fig. \ref{fig:FDGMBF},  Fig. \ref{fig:BAPlot},  Table \ref{tbl:det_mean}). The extraction fraction of FDG demonstrated to be close to that of Rb-chloride (Fig. \ref{fig:FDGext}). Hence, FDG $K_1$ quantification with the glucose-normalized EFC has the potential to provide MBF. 

The patient study is complex to do as it consists of both a dynamic FDG-PET scan and a dynamic Rb-PET scan, making patient accrual challenging. The sample size was small in this pilot study. \txtc{The range of the MBF values in this study was also limited with only one patient having a MBF$>$1.0 mL/min/cm$^3$. \txtb{The additive approach for glucose normalization seems slightly better than the multiplicative approach (Fig. \ref{fig:FDGMBF}, Fig. \ref{fig:FDGext}), but it remains open to further investigate the most appropriate normalization approach given the small sample size in the current study.} Another limitation is the use of $^{82}$Rb-chloride which is not an ideal tracer for MBF quantification. The results mainly provided a report to warrant future studies that should have a large sample size, include hyperaemic MBF values, and potentially use a better flow tracer (e.g. $^{11}$C-butanol) for comparison.} The analysis of this study was also limited to evaluation of global myocardial quantification instead of segment-level investigation to reduce the effect of noise. The noise performance of the scanner (2002 GE Discovery ST model) used in this study was far from optimal for exploring segment-based $K_1$, as indicated by the result from the previous identifiability analysis \cite{Zuo2020}. 

It is worth noting that the IDIF and myocardial TAC may suffer from the spill-over effect of a large ROI. Reducing the size of the left ventricle ROI can reduce the effect but it would increase the noise. The spill-over effect from the myocardium to the LV cavity mainly occurs in the late time of a FDG scan. It will not affect the FDG $K_1$ estimation much since the estimation of $K_1$ is more dominated by the early time points\cite{Zuo2019}. However, the spill-over may potentially lead to a bias in the $k_3$ and $k_4$ estimates, though the two kinetic parameters are not the main interest of the current study. Alternatively, the ascending aorta or arch of aorta may be used\txtb{, for example, for better quantification of myocardial metabolic rate of glucose (MRGlu) \cite{VW2001}}. One of our ongoing efforts is investigating the optimal option to extract IDIF for quantification of different myocardial FDG kinetic parameters \txtb{with an emphasis on FDG $K_1$}.

While motion correction was not included in the study, the effect of motion was less likely to result in a significant change to the results that were based on a large ROI given the spatial resolution of the PET scanner is only about 6-8 mm. A global myocardium ROI may involve the spill-over effect from the right ventricle. With the limited temporal sampling of 10 s/frame, the kinetic model used for FDG and Rb data in this study did not explicitly include the blood fraction from the right ventricle but only the left ventricle. The separate estimation of $v_{LV}$ and $v_{RV}$ parameters in \cite{Zuo2020} was therefore implicitly combined into a single blood volume parameter $v_b$ in Eq. (\ref{eq-2tct}). This is not uncommon. For example, Lortie {\em et al} also used a temporal sampling of 10 s/frame and single blood volume parameter for both $^{13}$N-ammonia and $^{82}$Rb studies. \txtc{We also tested the separate modeling of $v_{LV}$ and $v_{RV}$. The estimated $v_{RV}$ ($<4\%$) and its effect on $K_1$ were both small (results not shown). This is reasonable given that the ROI accounts for the whole myocardium which is less affected than a ROI that is solely nearby the right ventricle.}
    
Increase of temporal sampling, for example to 5 s/frame or 2 s/frame, has the potential to improve the separation of blood fractions from the left ventricle and right ventricle, and may lead to more robust estimation of FDG $K_1$. In particular, latest clinical PET scanners have an effective sensitivity gain of 4-25 fold and higher spatial resolution as compared to a typical conventional scanner GE Discovery 690 (see Table IV in \cite{Wang20TRPMS}), and are remarkably better than the GE DST scanner used in this study.  The EXPLORER total-body PET/CT scanner \cite{Cherry2017, Badawi2019} has an ultrahigh sensitivity for dynamic imaging, making it more feasible to explore higher temporal resolution and even pixel-wise parametric imaging in the myocardium. Furthermore, improved dynamic image reconstruction using machine learning concepts has been developed for dynamic PET imaging, e.g. with the kernel methods (e.g. \cite{Wang2015, Wang2019}) or deep neural-network methods (e.g. \cite{Gong2019, Reader2020}) as recent examples. Thus, the progress in PET instrumentation and algorithms may provide a future opportunity to exploit higher temporal resolution, higher spatial resolution, and also motion-corrected segment-based quantification for a better study design for enhanced evaluations of myocardial blood flow using these FDG methods. 

\section{Conclusion}

{This pilot study demonstrates that  FDG delivery rate $K_1$ from a one-hour dynamic scan was closely associated with MBF in the myocardium, especially after an adjustment for blood glucose levels. With the glucose normalization and extraction fraction correction that covert FDG $K_1$ to MBF, the FDG-derived MBF highly correlated with Rb MBF. The results also suggest FDG may have a first-pass myocardial extraction fraction similar to that of Rb-chloride. This work warrants a future, large study to further explore the potential of FDG for simultaneous imaging of myocardial blood flow and glucose metabolism.}

 \section*{Acknowledgments}

{The authors thank the anonymous reviewers for their very helpful review comments.} This work was supported in part by National Institutes of Health (NIH) under grant no. R21 HL131385 and American Heart Association under grant no. 15BGIA25780046. The work of J.E.L. is also supported in part by the Harold S. Geneen Charitable Trust Awards Program and the National Center for Advancing Translational Sciences, NIH, grant number UL1 TR001860 and linked award KL2 TR001859.
The authors thank Denise Caudle, Michael Rusnak, and Ben Spencer for their assistance in the dynamic PET/CT data acquisition, Diana Ramos for her efforts in patient recruitments, and the patients that agreed to participate in these studies.

\section*{{Appendix}}

\subsection*{Appendix I: Linearized relation between FDG $K_1$ and blood glucose level}
{The relation between FDG $K_1$ and blood flow $F$ can be described by the Renkin-Crone (RC) model without a specific consideration of capillary recruitment \cite{Carson2005},
\beq
K_1 = F\cdot\left[1-\exp\left(-\frac{PS}{F}\right)\right]\triangleq RC(F;PS),
\label{eq-rc}
\eeq
where $P$ denotes the permeability of FDG and $S$ is the surface area of a given section of the capillary bed. Here, the RC model is also denoted as a function of $F$ and $PS$ for convenient use. Following the classic Michaelis-Menten model, the product $PS$ relates to the blood glucose concentration $C_\glu$ following the form \txtc{(Eq. 11 of \cite{Carson2005}, Eq. 8 of \cite{Gjedde1980}, Eq. 2 of \cite{Gjedde1981})},
\beq
PS=\frac{V_{\mrm,\fdg}}{K_{\mrm,\fdg}+\frac{K_{\mrm,\fdg}}{K_{\mrm,\glu}}C_\glu}\triangleq \mm(C_\glu),
\label{eq-mm}
\eeq
where $V_{\mrm,\cdot}$ is the maximal rate of FDG (or glucose) transfer and $K_\mrm$ denotes the concentration of FDG (or glucose) at which the half-maximum transfer rate is reached.}

{Based on the above two models, FDG $K_1$ can be further expressed as a direct function of $C_\glu$, i.e.,
\beq
K_1\triangleq f(C_\glu)=RC\Big(F; \mm(C_\glu)\Big).
\label{eq-K1fun}
\eeq
By expanding $K_1$ at a reference blood glucose level $C_{\glu,0}$ using the first-order Taylor's expansion, we will have the following linear approximation,
\beq
K_1\approx f(C_{\glu,0})+\dot{f}(C_{\glu,0})\cdot\left(C_\glu-C_{\glu,0}\right),
\eeq
where $\dot{f}(C_{\glu,0})$ is the first-order derivative of $f$ at $C_{\glu,0}$,
\beq
\dot{f}(C_{\glu,0})=\left(\frac{\partial K_1}{\partial{PS}}\cdot\frac{\partial PS}{\partial{C_\glu}}\right)_{C_\glu=C_{\glu,0}}.
\eeq
From Eq. (\ref{eq-rc}), we have
\beq
\frac{\partial K_1}{\partial{PS}}=\frac{\partial RC(F;PS)}{\partial{PS}}=\exp\left(-\frac{PS}{F}\right)>0,
\label{eq-K1PS}
\eeq
and from Eq. (\ref{eq-mm}), we have
\beq
\frac{\partial PS}{\partial{C_\glu}}=\frac{\partial \mm(C_\glu)}{\partial{C_\glu}}=-\frac{V_{\mrm,\fdg}\frac{K_{\mrm,\fdg}}{K_{\mrm,\glu}}}{\left(K_{\mrm,\fdg}+\frac{K_{\mrm,\fdg}}{K_{\mrm,\glu}}C_\glu\right)^2}<0.
\eeq
Combing the above two relations together produces
\beq
\dot{f}(C_{\glu,0})<0.
\eeq}

Thus, the approximate linear relation of FDG $K_1$ with the blood glucose level $C_\glu$ can be re-expressed as
\beq
K_1\approx -s\cdot C_\glu + int,
\label{eq-K1C}
\eeq
where the absolute slope $s$ and the intercept $int$ relate to the reference blood glucose level $C_{\glu,0}$ via
\beq
s=-\dot{f}(C_{\glu,0})>0,\quad int = f(C_{\glu,0})+s\cdot C_{\glu,0}.
\eeq

\subsection*{Appendix II: Generalized RC model for FDG}

Following the hypothesis of Yoshida {\em et al.} \cite{Yoshida1996} for $^{82}$Rb and $^{13}$N-ammonia that capillary recruitment occurs at high coronary flows, we assume the PS product of FDG, is not a constant but depends on blood flow $F$ in a multicapillary system, 
\beq
\txtb{PS'= PS+\kappa\cdot F,}
\eeq
where \txtb{$PS'$ denotes the total PS product, $PS$ is the resting PS product, and $\kappa$} is a multiplicative coefficient. 

Substituting the above model into the original RC model in Eq. (\ref{eq-rc}), we then have the following generalized Renkin-Crone (GRC) model  for FDG,
\bea
K_1 &=& F\cdot\left[1-\exp\left(-\frac{PS+\kappa\cdot F}{F}\right)\right]\nonumber\\
&=&F\cdot\left[1-a_\fdg\cdot\exp\left(-\frac{PS}{F}\right)\right]\txtb{\triangleq GRC(F;PS)},
\eea
where 
\beq
a_\fdg=\exp(-\kappa)\txtb{>0}.
\label{eq-aFDG}
\eeq
\txtb{The GRC model is equal to Eq. (\ref{eq-rcfdg}) by setting $b_\fdg=PS$.} The coefficients $a_\fdg$ and $b_\fdg$ will be estimated by fitting the FDG $K_1$ and myocardial blood flow data.

\txtb{Note that with the GRC model, we can also derive the linearized relation between FDG $K_1$ and $C_\glu$ (Eq. (\ref{eq-K1C})) following the derivation in Appendix I for the RC model. The major change is that the term $\exp\left(-\frac{PS}{F}\right)$ in Eq. (\ref{eq-K1PS}) should be replaced by $a_\fdg\cdot\exp\left(-\frac{PS}{F}\right)$. The resulting $\frac{\partial K_1}{\partial{PS}}$ remains positive because $a_\fdg>0$ as shown in Eq. (\ref{eq-aFDG}).}

\section*{References} 
\bibliographystyle{unsrt}


\end{document}

%% file: myshortcuts.tex
\newcommand{\beq}{\begin{equation}}
\newcommand{\eeq}{\end{equation}}
\newcommand{\bea}{\begin{eqnarray}}
\newcommand{\eea}{\end{eqnarray}}
\newcommand{\ba}{\left(\!\!\begin{array}}
\newcommand{\ea}{\end{array}\!\!\right)}
\newcommand{\bc}{\begin{center}}
\newcommand{\ec}{\end{center}}

%% file: MBF_FDG_v8.bbl
\begin{thebibliography}{00}


\bibitem{Schindler2010} 
T. H. Schindler, H. R. Schelbert, A. Quercioli, V. Dilsizian,
``Cardiac PET Imaging for the Detection and Monitoring of Coronary Artery Disease and Microvascular Health," 
\emph{JACC: Cardiovascular Imaging}, 3(6): 623-640, 2010.

\bibitem{Kaufmann2005} 
P. A. Kaufmann, P. G. Camici,
``Myocardial Blood Flow Measurement by PET: Technical Aspects and Clinical Applications," 
\emph{Journal of Nuclear Medicine}, 46(1): 75-88, 2005.

\bibitem{DiCarli2007} 
MF Di Carli, S Dorbala, J Meserve, \emph{et al.},
``Clinical myocardial perfusion PET/CT," 
\emph{Journal of Nuclear Medicine}, 48(5): 783--793, 2007.

\bibitem{Murthy2018}
VL Murthy, TM Bateman, RS Beanlands, \emph{et al.},
``Clinical Quantification of Myocardial Blood Flow Using PET: Joint Position Paper of the SNMMI Cardiovascular Council and the ASNC," 
\emph{Journal of Nuclear Medicine}, 59(2):273-293, 2018.

\bibitem{Maddahi2014} 
J. Maddahi, R. R. S. Packard,
``Cardiac PET Perfusion Tracers: Current Status and Future Directions," 
\emph{Seminar in Nuclear Medicine}, 44(5): 333-43, 2014.

\bibitem{Iida1988} 
H Iida, I Kanno, A Takahashi, \emph{et al.},
``Measurement of absolute myocardial blood flow with H$_2^{15}$O and dynamic positron-emission tomography. Strategy for quantification in relation to the partial-volume effect," 
\emph{Circulation}, 78(1): 104–115, 1988.

\bibitem{Danad2014} 
I Danad, V Uusitalo, T Kero, \emph{et al.},
``Quantitative Assessment of Myocardial Perfusion in the Detection of Significant Coronary Artery Disease
Cutoff Values and Diagnostic Accuracy of Quantitative [15O]H2O PET Imaging," 
\emph{Journal of the American College of Cardiology}, 64(14): 1464-1475, 2014

\bibitem{Muzik1993} 
O. Muzik, RS Beanlands, GD Hutchins,  \emph{et al.},
``Validation of nitrogen-13-ammonia tracer kinetic model for quantification of myocardial blood flow using PET," 
\emph{Journal of Nuclear Medicine}, 34(1): 83-91, 1993.
		
\bibitem{Slomka2012} 
P J Slomka, E Alexanderson, R Jácome,  \emph{et al.},
``Comparison of Clinical Tools for Measurements of Regional Stress and Rest Myocardial Blood Flow Assessed with 13N-Ammonia PET/CT," 
\emph{Journal of Nuclear Medicine}, 53(2): 171-181, 2012.

\bibitem{Mullani1983} 
NA Mullani, RA Goldstein, KL Gould, \emph{et al.},
``Myocardial perfusion with rubidium-82. I. Measurement of extraction fraction and flow with external detectors," 
\emph{Journal of Nuclear Medicine}, 24(10): 898-906, 1983.

\bibitem{Lortie2007} 
M Lortie, RSB Beanlands, K Yoshinaga, \emph{et al.},
``Quantification of myocardial blood flow with $^{82}$Rb dynamic PET imaging," 
\emph{European journal of nuclear medicine and molecular imaging}, 34(11): 1765--1774, 2007.

\bibitem{Yoshida1996} 
K. Yoshida, N. Mullani, K. L. Gould, ``Coronary flow and flow reserve by PET simplified for clinical applications using rubidium-82 or nitrogen-13-ammonia,'' \emph{Journal of Nuclear Medicine}, 37(10):1701-1712., 1996.

\bibitem{ElFakhri2009} 
G El Fakhri, A Kardan, A Sitek, \emph{et al.},
``Reproducibility and Accuracy of Quantitative Myocardial Blood Flow Assessment with 82Rb PET: Comparison with 13N-Ammonia PET," 
\emph{Journal of Nuclear Medicine}, 50(7): 1062-1071, 2009.

\bibitem{Nesterov2014} 
SV Nesterov, E Deshayes, R Sciagrà, \emph{et al.},
``Quantification of Myocardial Blood Flow in Absolute Terms Using 82Rb PET Imaging: The RUBY-10 Study," 
\emph{JACC: Cardiovascular Imaging}, 7(11): 1119-1127, 2014.

\bibitem{Maddahi2012} 
J. Maddahi,
``Properties of an ideal PET perfusion tracer: new PET tracer cases and data," 
\emph{Journal of Nuclear Cardiology}, 19(suppl 1): S30-37, 2012.

\bibitem{vandenHoff2001} 
van den Hoff J, Burchert W, Borner AR,  \emph{et al.},
``1-C-11 Acetate as a quantitative perfusion tracer in myocardial PET," 
\emph{Journal of Nuclear Medicine}, 42(8):1174-82, 2001.

\bibitem{Sciacca2001} 
Sciacca RR, Akinboboye O, Chou RL, Epstein S, Bergmann SR. 
``Measurement of myocardial blood flow with PET using 1-C-11-acetate.," 
\emph{Journal of Nuclear Medicine}, 42(1):63-70, 2001.

\bibitem{Timmer2010} 
Timmer SAJ,  Lubberink M,  Germans T,  \emph{et al.},
``Potential of C-11 acetate for measuring myocardial blood flow: Studies in normal subjects and patients with hypertrophic cardiomyopathy," 
\emph{Journal of Nuclear Cardiology},  17(2):264-75, 2010.

\bibitem{Abraham2010} 
A Abraham, G Nichol, KA Williams,\emph{et al.},
``F-18-FDG PET Imaging of Myocardial Viability in an Experienced Center with Access to F-18-FDG and Integration with Clinical Management Teams: The Ottawa-FIVE Substudy of the PARR 2 Trial," 
\emph{Journal of Nuclear Medicine}, 51(4): 567-574, 2010.

\bibitem{Camici2008} 
P. G. Camici, S. K. Prasad, and O. E. Rimoldi,
``Stunning, hibernation, and assessment of myocardial viability," 
\emph{Circulation}, 117(1): 103-114, 2008.

\bibitem{Yama2003} 
H.Yamagishi, N. Shirai, M. Takagi, M. Yoshiyama, K. Akioka, K. Takeuchi , J. Yoshikawa.
``Identification of Cardiac Sarcoidosis with 13N-NH3/18F-FDG PET," 
\emph{Journal of Nuclear Medicine}, 44 (7) 1030-1036, 2003.
			
\bibitem{Maddahi2020} 
J. Maddahi, J. Lazewatsky, J. E. Udelson, \emph{et al.}, 
``Phase-III Clinical Trial of Fluorine-18 Flurpiridaz Positron Emission Tomography for Evaluation of Coronary Artery Disease," 
\emph{Journal of the American College of Cardiology}, 76(4): 391-401, 2020.

\bibitem{Carson2005} 
R. E. Carson,
``Tracer Kinetic Modeling in PET," 
in \emph{Positron Emission Tomography}, D. L. Bailey, D. W. Townsend, P. E. Valk, and M. N. Maisey Eds. Springer London, pp 127-159, 2005.

\bibitem{Tseng2004} 
J Tseng, LK Dunnwald, EK Schubert, \emph{et al.}, 
``Correlation with tumor blood flow and changes in response to neoadjuvant chemotherapy," 
\emph{Journal of Nuclear Medicine}, 45(11): 1829-1837, 2004.

\bibitem{Mullani2008} 
N. A. Mullani, R. S. Herbst, R. G. O'Neil, \emph{et al.}, 
``Tumor blood flow measured by PET dynamic imaging of first-pass F-18-FDG uptake: A comparison with O-15-Labeled water-measured blood flow," 
\emph{Journal of Nuclear Medicine}, 49(4): 517-523, 2008.

\bibitem{Bernstine2011} 
H. Bernstine,  M Braun, N Yefremov, \emph{et al.}, 
``FDG PET/CT Early Dynamic Blood Flow and Late Standardized Uptake Value Determination in Hepatocellular Carcinoma," 
\emph{Radiology}, 260(2): 503-510, 2011.

\bibitem{Cochet2012} 
A Cochet, S Pigeonnat, B Khoury, \emph{et al.}, 
``Evaluation of Breast Tumor Blood Flow with Dynamic First-Pass F-18-FDG PET/CT: Comparison with Angiogenesis Markers and Prognostic Factors," 
\emph{Journal of Nuclear Medicine}, 53(4): 512-520, 2012.

\bibitem{Humbert2018} 
O Humbert, M Lasserre, A Bertaut,  \emph{et al.}, 
``Breast Cancer Blood Flow and Metabolism on Dual-Acquisition F-18-FDG PET: Correlation with Tumor Phenotype and Neoadjuvant Chemotherapy Response," 
\emph{Journal of Nuclear Medicine}, 59(7): 1035-1041, 2018.

\bibitem{Winterdahl2011} 
M. Winterdahl, O. L. Munk, M. Sorensen, F. V. Mortensen, and S. Keiding.
``Hepatic Blood Perfusion Measured by 3-Minute Dynamic F-18-FDG PET in Pigs," 
\emph{Journal of Nuclear Medicine}, 52(7): 1119-1124, 2011.

\bibitem{Walberer2012} 
M. Walberer, H Backes, MA Rueger,  \emph{et al.}, 
``Potential of Early F-18 -2-Fluoro-2-Deoxy-D-Glucose Positron Emission Tomography for Identifying Hypoperfusion and Predicting Fate of Tissue in a Rat Embolic Stroke Model," 
\emph{Stoke}, 43(1): 193-198, 2012.

\bibitem{Zuo2020} 
Y. Zuo, RD Badawi, C. Foster, T. Smith, J. Lopez, and G. Wang,
``Multiparametric cardiac 18F-FDG PET in humans: kinetic model selection and identifiability analysis," 
\emph{IEEE Transactions on Radiation in Plasma and Medical Sciences}, 4(6): 759-767, 2020.

\bibitem{Gambhir1989} 
S S Gambhir, M Schwaiger, S C Huang, J Krivokapich, H R Schelbert, C A Nienaber, M E Phelps,
``Simple noninvasive quantification method for measuring myocardial glucose utilization in humans employing positron emission tomography and fluorine-18 deoxyglucose," 
\emph{J Nucl Med}, 30(3): 359-66., 1989.

\bibitem{Naganawa2020} 
M Naganawa, JD Gallezot, V Shah, T Mulnix, C Young, M Dias, MK Chen, AM Smith, RE Carson,
``Assessment of population-based input functions for Patlak imaging of whole body dynamic 18 F-FDG PET," 
\emph{EJNMMI Physics}, 7(1):67, 2020.

\bibitem{Bland1986} 
J. M. Bland and D. G. Altman,
``Statistical methods for assessing agreement between two methods of clinical measurement," 
\emph{Lancet}, 1(8476): 307-310, 1986. 

\bibitem{Cerqueira2002} 
M. D. Cerqueira, N. J. Weissman, V. Dilsizian,  \emph{et al.}, 
``Standardized myocardial segmentation and nomenclature for tomographic imaging of the heart - A statement for healthcare professionals from the Cardiac Imaging Committee of the Council on Clinical Cardiology of the American Heart Association," 
\emph{Circulation}, vol. 105, no. 4, pp. 539-542, Jan 2002, 

\bibitem{Innis2007} 
R.B. Innis, V.J. Cunningham, J. Delforge, \emph{et al.}, 
``Consensus nomenclature for in vivo imaging of reversibly binding radioligands," 
\emph{J Cereb Blood Flow Metab }, 27: 1533- 1539, 2007.

\bibitem{Stout1998} 
D. B. Stout, S. C. Huang, W. P. Melega, M. J. Raleigh, M. .E Phelps, J. R. Barrio,
``Effects of large neutral amino acid concentrations on 6-[F-18]Fluoro-L-DOPA kinetics," 
\emph{J Cereb Blood Flow Metab}, 18(1):43-51, 1998.

\bibitem{Marshall1998} 
R C Marshall, P Powers-Risius, R H Huesman, B W Reutter, S E Taylor, H E Maurer, M K Huesman, T F Budinger,
``Estimating glucose metabolism using glucose analogs and two tracer kinetic models in isolated rabbit heart," 
\emph{JAm J Physiol}, 275(2):H668-79, 1998.

\bibitem{Zuo2019} 
Zuo Y, Sarkar S, Corwin MT, Olson K, Badawi RD, Wang GB.
``Structural and practical identifiability of dual-input kinetic modeling in dynamic PET of liver inflammation," 
\emph{Physics in Medicine and Biology}, 64(17): 175023 (18pp),  2019.

\bibitem{VW2001} 
Van der Weerdt AP, Klein LJ, Boellaard R, Visser CA, Visser FC, Lammertsma AA.
``Image-derived input functions for determination of MRGlu in cardiac 18F-FDG PET scans," 
\emph{J Nucl Med}, 42: 1622-1629, 2001.

\bibitem{Wang20TRPMS} 
G. Wang, A. Rahmim, R. N. Gunn,
``PET parametric imaging: past, present, and future," 
\emph{IEEE Transactions on Radiation and Plasma Medical Sciences}, 4(6): 663 - 675, 2020. 

\bibitem{Cherry2017} 
S. Cherry, R. Badawi, J. Karp, W. Moses, P. Price, and T. Jones,
``Total-body imaging: Transforming the role of PET in translational medicine," 
\emph{Science Translational Medicine}, vol. 9, no. 381, p. eaaf6169, 2017.

\bibitem{Badawi2019} 
R. Badawi, H Shi, P Hu, \emph{et al.}, 
``First human imaging studies with the EXPLORER total-body PET scanner," 
\emph{Journal of Nuclear Medicine}, vol. 60, no. 3, pp. 299-303, 2019.

\bibitem{Wang2015} 
G. Wang, J. Qi, 
\newblock ``PET image reconstruction using kernel method,'' 
\newblock {\em IEEE Transactions on Medical Imaging}, vol. 34, no. 1, pp. 61-71, 2015.

\bibitem{Wang2019} 
G. Wang,
\newblock ``High temporal-resolution dynamic PET image reconstruction using a new spatiotemporal kernel method,'' 
\newblock {\em IEEE Transactions on Medical Imaging}, vol. 38, no. 3, pp. 664 – 674, 2019.

\bibitem{Reader2020} 
A. Reader, G. Corda, A. Mehranian, C. da Costa-Luis, S. Ellis, and J. Schnabel,
``Deep Learning for PET Image Reconstruction," 
\emph{IEEE Transactions on Radiation and Plasma Medical Sciences}, 5(1): 1-25, 2020.

\bibitem{Gong2019} 
K. Gong, J. Guan, K. Kim, {\em et. al.}, ``Iterative PET Image Reconstruction Using Convolutional Neural Network Representation,'' \emph{IEEE Transactions on Medical Imaging}, vol. 38, no. 3, pp. 675-685, 2019.

\bibitem{Gjedde1980} 
A. Gjedde, ``Rapid steady-state analysis of blood-brain glucose transfer in rat,'' \emph{Acta Physiol Scand}, 108(4):331-339, 1980.

\bibitem{Gjedde1981} 
A. Gjedde, ``High- and low-affinity transport of D-glucose from blood to brain,'' \emph{Journal of Neurochemistry}, 36(4):1463-71, 1981.




\end{thebibliography}
